\documentstyle[nato,epsf]{crckapb} 

\def \deg    {$^{\circ}$}
\def \etal   {{et al.\thinspace}}
\def \eg     {{e.g.,}}

\def \ie     {{i.e.,}}

\def\gtorder	{\mathrel{\raise.3ex\hbox{$>$}\mkern-14mu\lower0.6ex\hbox{$\sim$}}}
\def\ltorder	{\mathrel{\raise.3ex\hbox{$<$}\mkern-14mu\lower0.6ex\hbox{$\sim$}}}

\def\lsim	{ \rlap{\lower .5ex \hbox{$\sim$} }{\raise .4ex \hbox{$<$} } }
\def\gsim	{ \rlap{\lower .5ex \hbox{$\sim$} }{\raise .4ex \hbox{$>$} } }
\def\solar	{ {\odot} }

\def\msolar	{ \rm {M_{\odot}} }
\def\earth	{ {\oplus} }

\def\mearth	{ \rm {M_{\oplus}} }

\def\etal	{{\it et~al.}}

\def\abs	{ \hbox{ \vrule height .8em depth .4em width .6pt } \,} 
\def\apj	{{\it Astrophys. J.\/}}
\def\mnras	{{\it Mon. Not. Royal Astron. Soc.\/}}
\def\pasp	{{\it Proc. Astron. Soc. Pacific\/}}
\def\aap	{{\it Astron. Astrophysics\/}}
\begin{opening}
\title{Searching for Unseen Planets via\protect\\
       Occultation and Microlensing}

\author{Penny D. Sackett}
\institute{Kapteyn Astronomical Institute\\
           Postbus 800, 9700 AV Groningen, The Netherlands\\
           psackett@astro.rug.nl}

\end{opening}

\runningtitle{OCCULTATION AND MICROLENSING PLANET SEARCHES}

\begin{document}


\begin{abstract}

The fields of occultation and microlensing are linked historically. 
Early this century, occultation of the Sun by the Moon allowed the apparent 
positions of background stars projected near the limb of the Sun to be 
measured and compared with their positions six months later 
when the Sun no longer influenced the light path to Earth.  
The measured shift in the stellar positions was consistent with lensing 
by the gravitational field of the Sun during the occultation,  
as predicted by the theory of general relativity. This series of 
lectures explores the principles, possibilities and challenges 
associated with using occultation and microlensing to discover and 
characterize unseen planets orbiting distant stars.  
The two techniques are complementary in terms of the information 
that they provide about planetary systems and the 
range of system parameters to which they are most sensitive.  
Although the challenges are large, both microlensing and occultation 
may provide avenues for the discovery of extra-solar planets as small 
as Earth.

\end{abstract}


\section{Introduction}

Indirect methods to search for extra-solar planets do not measure  
emission from the planet itself, but instead seek to discover and 
quantify the tell-tale effects that the planet would have on the 
position (astrometry) and motion (radial velocity) of its parent star, 
or on the apparent brightness of its parent star (occultation) 
or random background sources (gravitational microlensing).  
All of these indirect signals have a characteristic temporal behavior 
that aids in the discrimination between planetary effects and other 
astrophysical causes.  
The variability can be due to the changing position of the planet with 
respect to the parent star (astrometry, radial velocity, occultation), 
or the changing position of the complete planetary system with respect 
to background stars (microlensing).  
The time-variable photometric signals that can be measured using 
occultation and microlensing techniques are the focus of this small series 
of lectures.   

An occultation is the temporary dimming of the apparent brightness of 
a parent star that occurs when a planet transits the stellar disk;  
this can occur only when the orbital plane is nearly perpendicular 
to the plane of the sky. 
Because the planet is considerably cooler than its parent star, 
its surface brightness at optical and infrared wavelengths is less, 
causing a dip in the stellar light curve whenever the planet 
(partially) eclipses the star.  
Since the fractional change in brightness is proportional to the fraction of 
the stellar surface subtended by the planetary disk,   
photometric measurements directly yield a measure of the planet's size.
For small terrestrial planets, the effect is simply to occult a fraction 
of the stellar light; the atmospheres of larger gaseous planets 
may also cause absorption features that can be measured during transit with   
high resolution, very high S/N spectroscopic monitoring.  

The duration of a transit is a function of the size of the stellar disk 
and the size and inclination of the planetary orbit.  
Together with an accurate stellar typing of the parent star, measurement 
of the transit duration and period provides an estimate for the 
radius and inclination of the planet's orbital plane. 
Since large planets 
in tight orbits will create the most significant and frequent 
occultations, these are the easiest to detect.  
If hundreds of stars can be monitored with significantly better 
than 1\% photometry, the transit method can be applied from the 
ground to place statistics on Jupiter-mass planets in tight orbits.
Space-based missions, which could search for transits continuously and with 
higher photometric precision, may be capable of detecting 
Earth-mass planets in Earth-like environments via the occultation method.  
Moons or multiple planets may also be detectable, not through their eclipsing 
effect, but by the periodic change they induce in the timing of 
successive transits of the primary occulting body. 

Microlensing occurs when a foreground compact object (\eg\ a star, perhaps  
with orbiting planets) moves between an observer and a luminous 
background source (\eg\ another star).  The gravitational field of 
the foreground lens alters the path of the light from the background source, 
creating multiple images with a combined brightness larger than that of 
the unlensed background source.  
For stellar or planetary mass lenses, the separation 
of these images is too small to be resolved, but the combined brightness 
of the images changes with time in a predictable manner as the lensing 
system moves across the sky with respect to the background source.  
Hundreds of microlensing events have been detected in the Galaxy, 
a large fraction of which are due to (unseen) stellar lenses.  
In binary lenses with favorable geometric configurations, the 
lensing effect of the two lenses combines in a 
non-linear way to create detectable and rapid variations 
in the light curve of the background source star.  Modeling of these 
features yields estimates for the mass ratio and 
normalized projected orbital radius for the binary lens; 
in general, smaller-mass companions produce weaker and shorter deviations.

Frequent, high-precision 
photometric monitoring of microlensing events 
can thus be used to discover and characterize 
extreme mass-ratio binaries (\ie\ planetary systems).   
With current ground-based technology, 
microlensing is particularly suited to the 
detection of Jupiter-mass planets in Jupiter-like environments.
Planets smaller than Neptune will resolve the brightest background sources 
(giants) diluting the planetary signal. 
For planets above this mass, the planetary detection 
efficiency of microlensing is a weak function of the planet's mass 
and includes a rather broad range in orbital radii, making it 
one of the best techniques for a statistical study of the frequency 
and nature of planetary systems in the Galaxy.  
Microlensing can discover planetary systems at distances of the 
Galactic center and is the only technique that is capable of 
detecting unseen planets around {\it unseen parent stars\/}!

These lectures begin with a discussion of the physical basis of 
occultation and microlensing, emphasizing their strengths and weaknesses 
as well as the selection effects and challenges 
presented by sources of confusion for the planetary signal.  
The techniques are then placed in the larger 
context of extra-solar planet detection.  Speculative comments about possibilities in the next decade cap the 
lectures.


\section{Principles of Planet Detection via Occultations}

Due to their small sizes and low effective temperatures, planets 
are difficult to detect directly.  Compared to stars, their  
luminosities are reduced by the square of the ratio of their radii 
(factors of $\sim$$10^{-2} - 10^{-6}$ in 
the Solar System) and the 
fourth power of the ratio of their effective temperatures 
(factors of $\sim$$10^{-4} - 10^{-9}$ in the Solar System). 
Such planets may be detected indirectly however if they chance 
to transit (as viewed by the observer) the face of their parent star 
and are large enough to occult a sufficient fraction of 
the star's flux.   This method of detecting planets around 
other stars was discussed as early as mid-century (Sturve 1952), 
but received serious attention only after the detailed 
quantification of its possibilities by Rosenblatt (1971) and 
Borucki and Summers (1984).

Such occultation effects have been observed for many years in 
the photometry of binary star systems whose orbital planes 
lie close enough to edge-on as viewed from Earth that the disk of 
each partner occults the other at some point during the orbit, creating 
two dips in the combined light curve of the system.  
The depth of the observed occultation depends on the 
relative size and temperatures of the stars.   
For planetary systems, only the dip caused by the occultation of the brighter 
parent star by the transit of the smaller, cooler planet will be detectable. 
The detection rate for a given planetary system will depend on 
several factors: the geometric probability that a transit will occur, 
the frequency and duration of the observations compared to the 
frequency and duration of the transit, and the sensitivity of the 
photometric measurements compared to the fractional deviation in 
the apparent magnitude of the parent star due to the planetary 
occultation.   We consider each of these in turn.

\hglue 1cm
\epsfxsize=9cm\epsffile{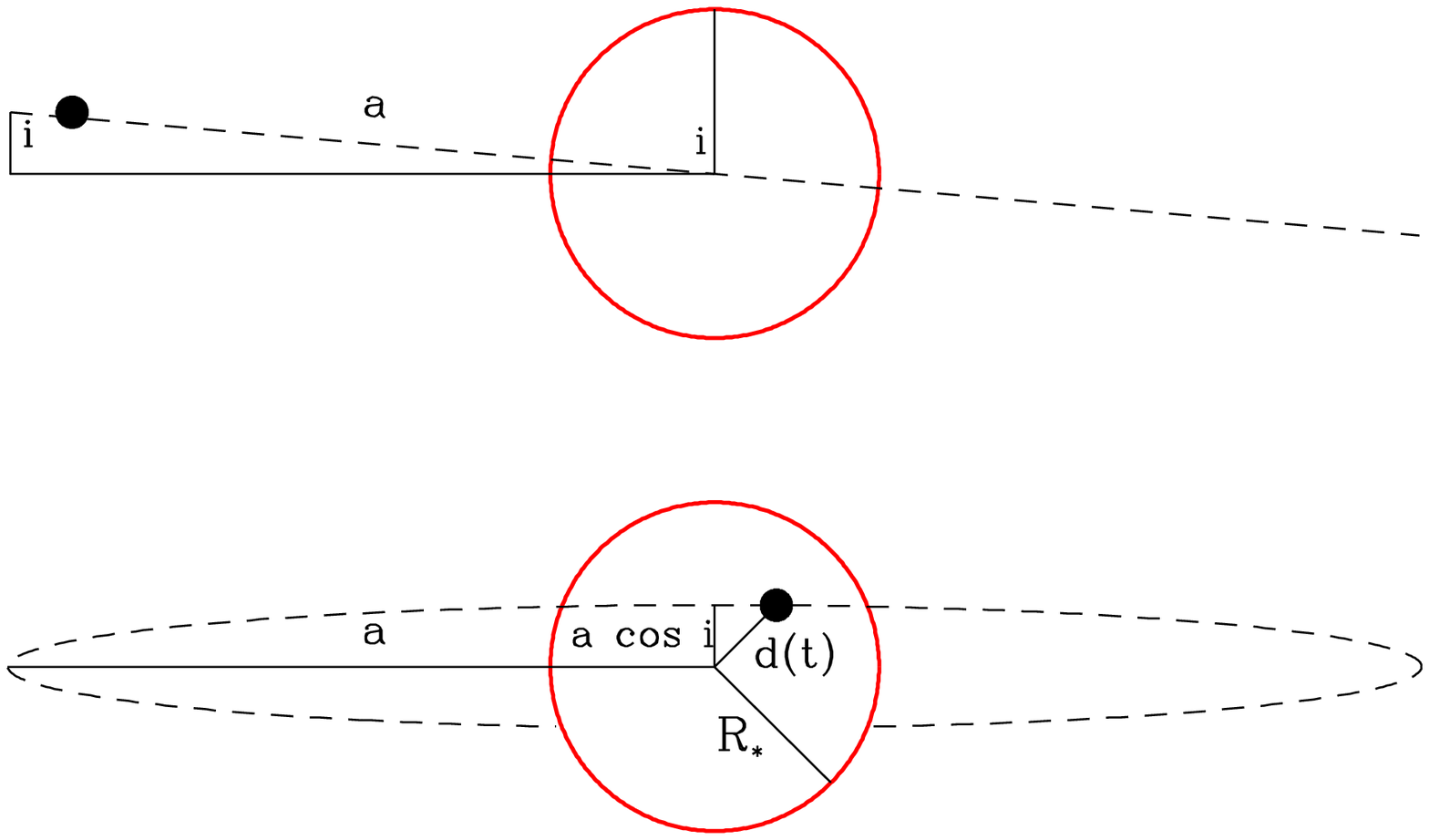}

\vglue -3.55cm 
{\small Fig.~1 --- Geometry of a transit 
event of inclination $i$ and orbital radius $a$ 
as seen from the side (top) and observer's vantage point (bottom) at a moment 
when the planet lies a projected distance $d(t)$ from the stellar center.} 
\vskip 0.4cm

Unless stated otherwise in special cases below, 
we will assume for the purposes of 
discussion that planetary orbits are circular and 
that the surface brightness, mass, and radius of the planet 
are small compared to that of the parent star.  
We will also assume that the orbital radius is much larger than 
the size of the parent star itself.  

\subsection{Geometric Probability of a Transit}

Consider a planet of radius $R_p$ orbiting a star of radius $R_*$ 
and mass $M_*$ at an orbital radius $a$.  A transit of the stellar 
disk will be seen by an external observer only if the orbital 
plane is sufficiently inclined with respect to the sky plane (Fig.~1). 
In particular, the inclination $i$ must satisfy

\begin{equation}
a \, \cos{i} \leq R_* + R_p~~.
\end{equation}

Since $\cos{i}$ is simply the projection of the 
normal vector (of the orbital plane) onto the sky plane, it is 
equally likely to take on any random value between 0 and 1.  
Thus, for an ensemble of planetary systems with arbitrary 
orientation with respect to the observer, the probability 
that the inclination satisfies the geometric criterion for a transit 
is:

\begin{equation}
{\rm Geometric \, \, Transit \, \, Prob} =  
\frac{ \int _{_0} ^{(R_* + R_p)/a} \, d(\cos{i}) }
{ \int _{_0} ^{^1} \, d(\cos{i}) } 
= \frac{R_* + R_p}{a} \approx \frac{R_*}{a}
\end{equation}

Geometrically speaking, the occultation method favors those 
planets with small orbital radii in systems with large parent stars. 
As can be seen in Fig.~2, for planetary systems like the Solar System 
this probability is small:  $\ltorder 1\%$ for 
inner terrestrial planets and about a factor of 10 smaller for 
jovian gas giants.  This means that unless a method can be 
found to pre-select stars with ecliptic planes oriented perpendicular 
to the plane of the sky, thousands of random stars must 
be monitored in order to detect statistically meaningful numbers of 
planetary transits due to solar systems like our own.

\hglue 1cm
\epsfxsize=10cm\epsffile{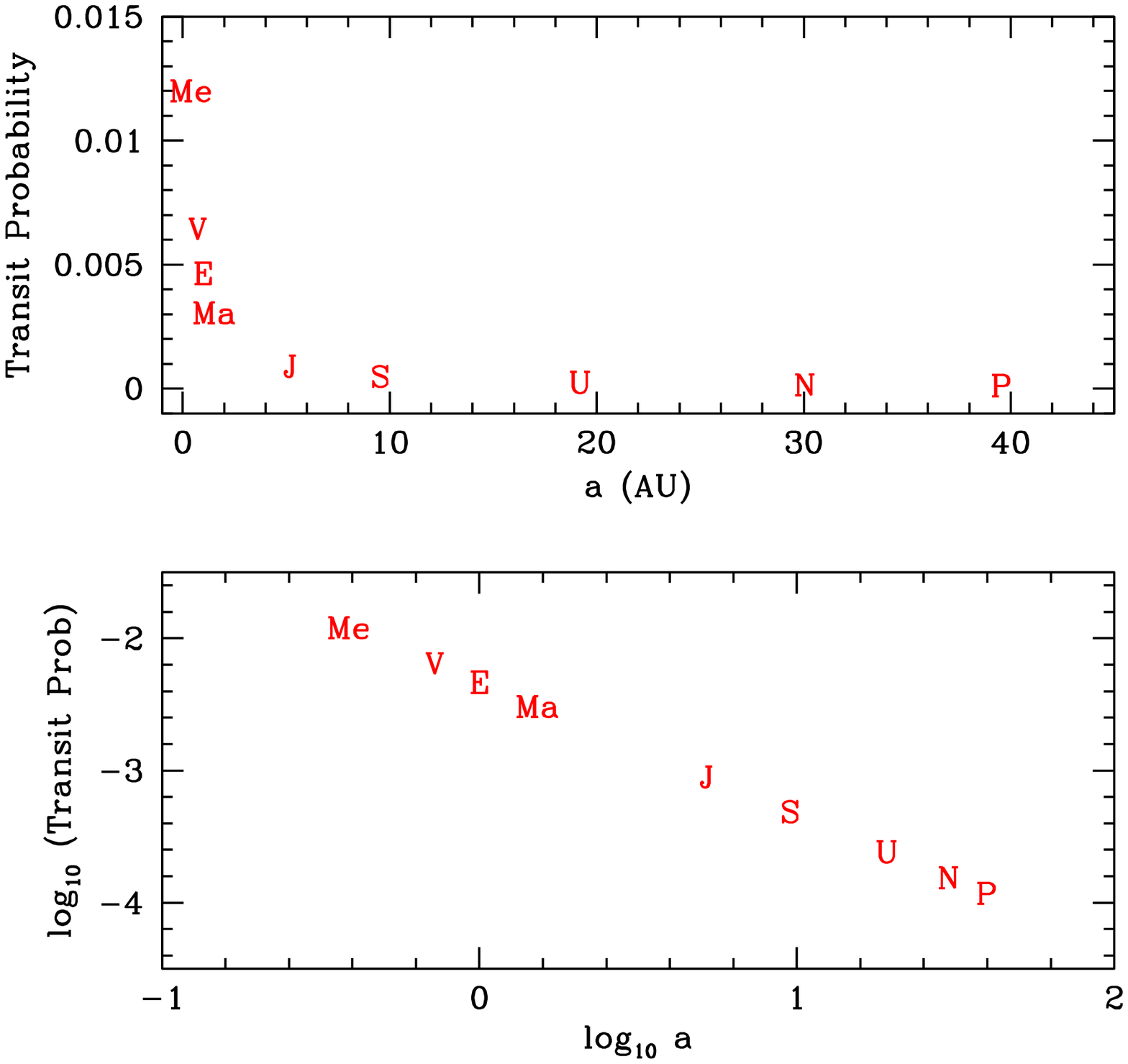}

\vglue -0.5cm
{\small Fig.~2 --- 
Probability of transits by Solar System objects 
as seen by a random external observer.}
\vskip 0.4cm

\subsubsection{Inclination Pre-selection} 

Under the assumption that the orbital angular momentum vector of a planetary 
system and the rotational angular momentum vector of the parent star 
share a common origin and thus a common direction, single stars can 
be pre-selected for transit monitoring programs on the basis of a 
measurement of their rotational spin.  In this way, one may hope 
to increase the chances of viewing the planetary orbits edge-on.  
Through spectroscopy, the line-of-sight component of the  
rotational velocity $v_{*, \, los}$ of a star's atmosphere can be measured.  
The period $P_{*, \, rot}$ of the rotation 
can be estimated by measuring the periodic photometric signals 
caused by sunspots, and the radius $R_*$ of the star can be determined 
through spectral typing and stellar models.  An estimate for the 
inclination of the stellar rotation plane to the plane of the sky 
can then be made:
\begin{equation}
\sin{i_{*, \, rot}} =\frac{v_{*, \, los} \, \, P_{*, \, rot}}{2\pi \, R_*} ~~~,
\end{equation}

\noindent
and only those stars with high rotational inclinations selected to 
be monitored for transits.  

How much are the probabilities increased by such pre-selection?  
Fig.~3 shows the probability 
of the planetary orbital inclination being larger (more edge-on) than 
a particular value ranging from 89.5\deg\ $< i < 85$\deg, 
if the parent star is pre-selected to have a rotational plane 
with inclination $i_{*, \, rot} \ge i_{\rm select}$.

\hglue 1cm
\epsfxsize=9cm\epsffile{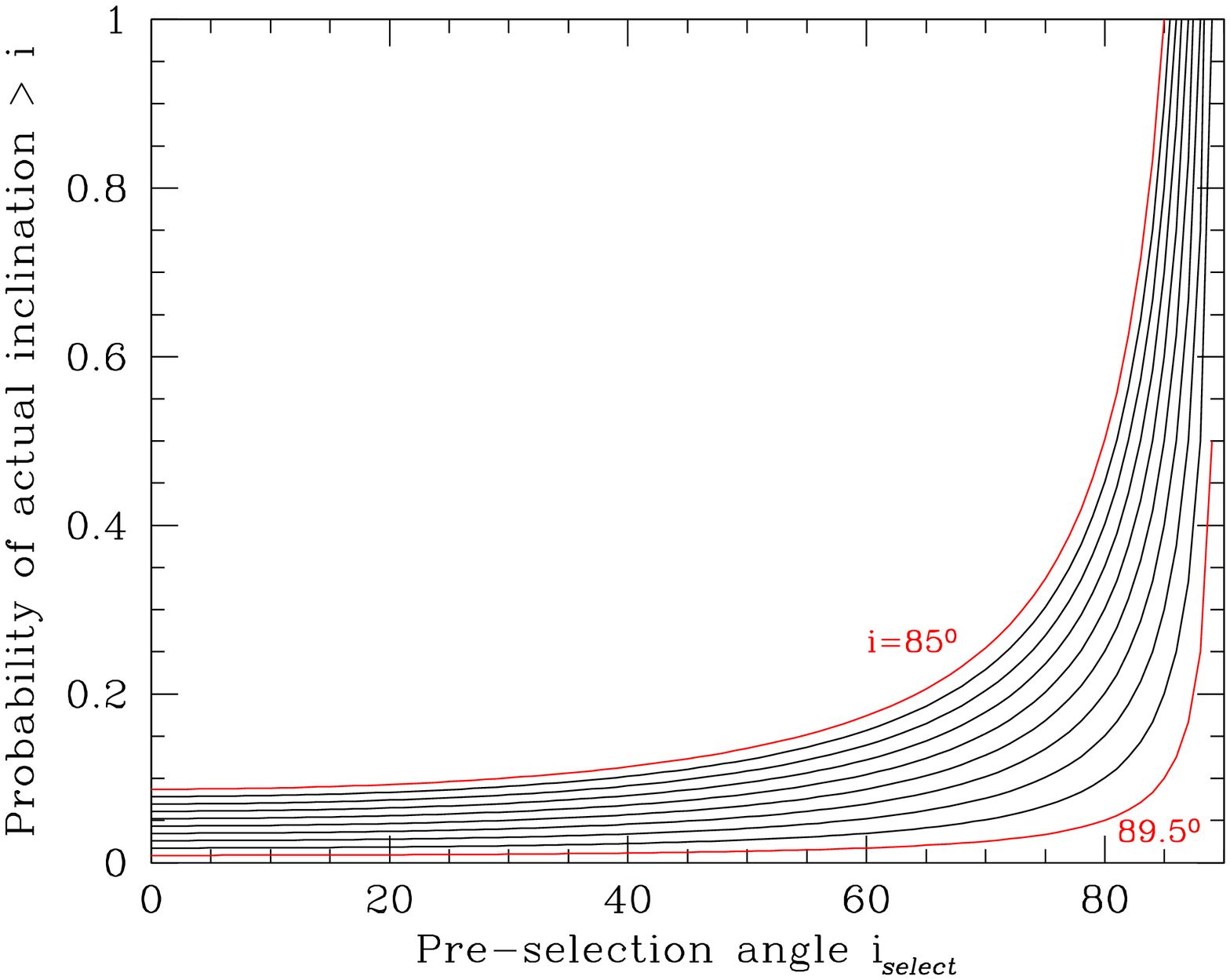}

\vglue -1.5cm{\small Fig.~3 --- Increase of geometric transit probability 
through pre-selection of the inclination angle to be larger than 
$i_{\rm select}$, for example through measurement of the rotational 
spin of the parent.}
\vskip 0.4cm

In order to produce a detectable transit, most planets will require 
an orbital inclination $\ltorder 1$\deg\ from edge-on.  
If planetary systems could be pre-selected to have $i > 85$, 
the geometric transit probability would be increased by a factor of $\sim$10.  
Unfortunately, measurement uncertainties in the quantities 
required to determine $\sin{i_{*, \, rot}}$ are likely to remove much  
of the advantage that pre-selection would otherwise afford. 
Since $\delta(\cos{i}) = - \tan{i} \, \delta(\sin{i})$,  
even small errors in $\sin{i_{*, \, rot}}$ translate into large 
uncertainties in $\cos{i_{*, \, rot}}$ and thus the probability 
that a transit will occur.  
Furthermore, an accurate measurement of 
$\cos{i_{*, \, rot}}$ does not ensure that $\cos{i}$ for the 
planetary orbital plane is known.  The planets in our 
own Solar System are misaligned by about 7\deg\ with the Sun's rotational 
plane, a result that is similar to that found for binaries orbiting 
solar-type stars (Hale 1994).  It is thus reasonable 
to assume that an accurate measurement of $i_{*, \, rot}$ will constrain  
the planetary orbital plane only to within $\sim$10\deg.  

To enhance probabilities, current ground-based attempts 
to detect transits have taken a different tack by 
concentrating on known eclipsing binary star systems in which the orbital 
plane of the binary is known to be close to edge-on.   
Assuming that any other companions will have similarly aligned angular momentum 
vectors, it is hoped that such systems will have larger than 
random chances of producing a transit event.  
The precession of orbital plane likely to be present in such systems 
may actually bring the planet across the face of the star more 
often than in single star systems (Schneider 1994). 
On the other hand, the evolution 
and dynamics of single and double star systems is so 
different that the formation and frequency of their planetary companions 
is likely to be quite different as well.  In particular, it may be difficult 
for planets in some binary systems to maintain long-lived circular  
orbits and thus, perhaps, to become the birth place 
of life of the sort that has evolved on Earth.   

Given the uncertainties involved, inclination pre-selection 
in single stars is unlikely to increase geometric transit 
probabilities by factors larger than 3 -- 5.  
Ambitious ground-based and space-based initiatives, however, 
may monitor so many stars that pre-selection is not necessary.

\subsection{Transit Duration}

The duration and frequency of the expected transits will determine the 
observational strategy of an occultation program.  The frequency is 
simply equal to one over the orbital period $P = \sqrt{4 \pi^2 a^3 / G M_*}$.
If two or more transits for a given system can be measured and 
confirmed to be due to the same planet, the period $P$ and 
orbital radius $a$ are determined.  
In principle, the ratio of the transit duration to the total 
duration can then be used to determine the inclination of the orbital 
plane, if the stellar radius is known.    

The duration of the transit will be equal to the fraction 
of the orbital period $P$ during which the 
projected distance $d$ between the centers of the star and 
planet is less than the sum of their radii $R_* + R_p$.  
Refering to Fig.~4 we have 
\begin{equation}
{\rm Duration} \equiv t_T = \frac{2 \, P}{2 \pi} \arcsin
{\left( \frac{\sqrt{(R_* + R_p)^2 - a^2 \cos^2{i}}}{a} \right)} ~~~,
\end{equation}

\noindent 
which for $a >> R_* >> R_p$ becomes 

\begin{equation}
t_T = \frac{P}{\pi} \sqrt{\left(\frac{R_*}{a}\right)^2 - \cos^2{i}} \ \
\leq \ \ \frac{P \, R_*}{\pi \, a} ~~~.
\end{equation}

\noindent 
Note that because the definition of a transit requires 
that $a\cos{i} \leq (R_* + R_p)$, 
the quantity under the square root in Eq.~4 does not become negative.   

\vskip -2.25cm
\hglue -1cm
\epsfxsize=13.5cm\epsffile{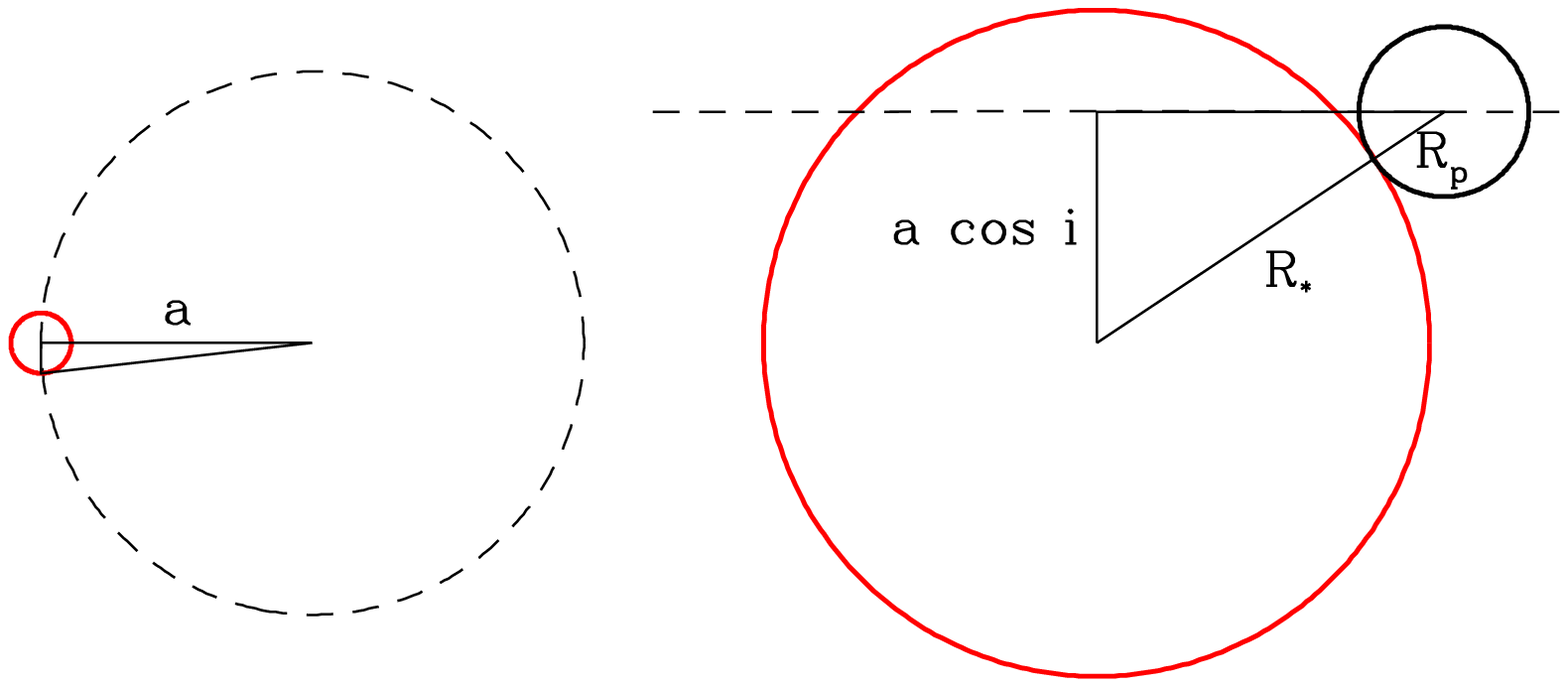}

\vglue -6.25cm{\small Fig.~4 --- Transit duration is set by  
fraction of total orbit (left) for which a portion of the planet eclipses the stellar disk (right).}
\vskip 0.4cm

Fig.~5 shows the maximum 
transit duration and period for planets in the Solar System. 
In order to confirm a planetary detection with one or more additional  
transits after the discovery of the first eclipse, a 5-year 
experiment can be sensitive to planets orbiting solar-type stars only 
if their orbital radius is equal to or smaller than that of Mars. 
Such planets will have transit durations of less than one day, requiring 
rapid and continuous sampling to ensure high detection probabilities. 

The actual transit duration depends sensitively 
on the inclination of the planetary 
orbit with respect to the observer, as shown in Fig.~6.  
The transit time of Earth as seen by an external observer changes from 
0.5 days to zero (no transit) if the observers viewing angle is more 
than 0.3\deg\ from optimal.  Since the orbital planes of any two of the 
inner terrestrial planets in the  
\newpage 

\vglue -1.15cm
\hglue -0.5cm
\epsfxsize=12cm\epsffile{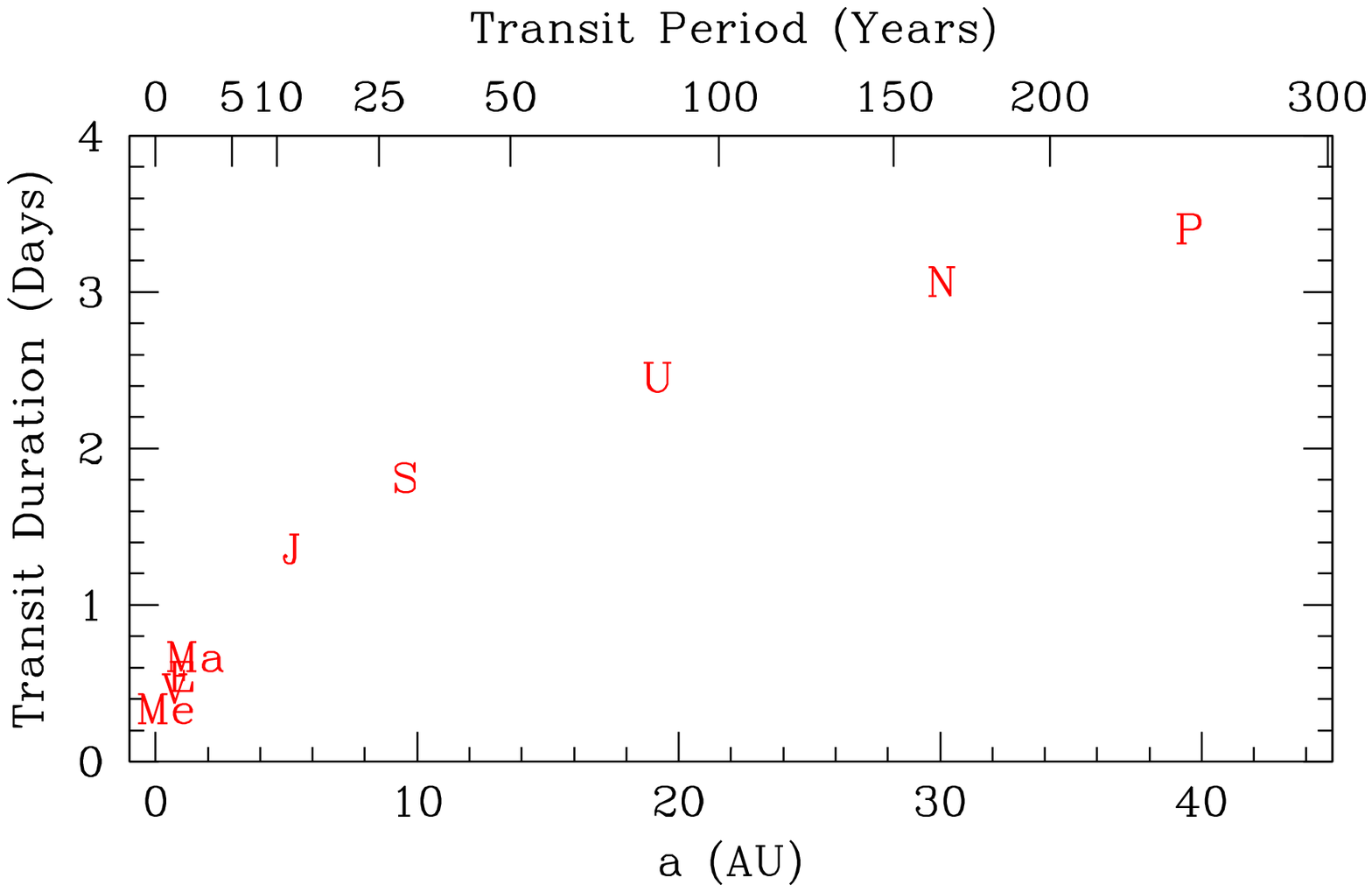}

\vskip -4.6cm {\small Fig.~5 --- Edge-on transit durations and 
periods for Solar System planets.}
\vskip 0.4cm

\noindent 
Solar System are misaligned 
by 1.5\deg\ or more, if other planetary 
systems are like our own, a given observer would expect to see transits 
from only one of the inner planets.  
This would decrease the detection probabilities for 
planetary systems, but also the decrease the probability of 
incorrectly attributing transits from different planets to 
successive transits of one (mythical) shorter period object.  

\vglue -0.5cm
\hglue 1.5cm
\epsfxsize=8cm\epsffile{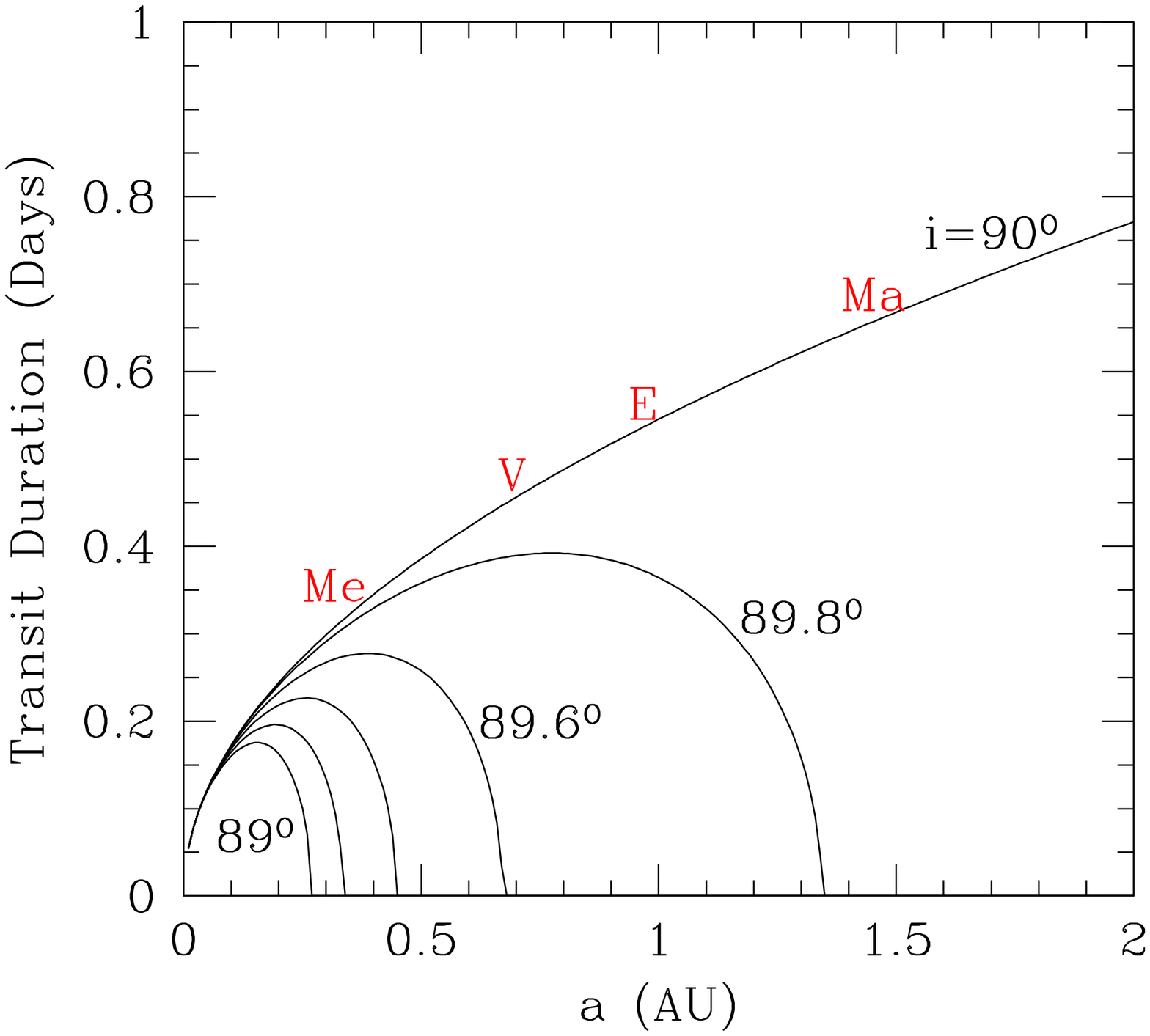}

\vglue -0.3cm{\small Fig.~6 --- ``Inner planet'' transit durations 
for different inclinations ($R_* = R_\odot$).}
\vskip 0.4cm

If the parent star can be typed spectroscopically, stellar models 
can provide an estimate for the stellar radius $R_*$ in the 
waveband in which the photometric partial eclipse was 
measured.  (It is important to match wavebands 
since limb-darkening can make the star look larger at redder  
wavelengths which are more sensitive to the cooler 
outer atmosphere of the star.)  The temporal resolution of a single 
transit then places a lower limit on the orbital radius $a$ of the 
planet, but a full determination of $a$ requires 
knowledge of the period from multiple transit 
timings which remove the degeneracy due to the otherwise unknown 
orbital inclination.  In principle, if the limb darkening of 
the parent star is sufficiently well-understood, measurements in 
multiple wavebands can allow an estimate for the inclination, and 
thus for $a$ from a single transit; this is discussed more 
fully in \S2.3.1.

\subsection{Amplitude and Shape of the Photometric Signature} 

Planets with orbital radii of 2~AU or less orbiting stars even 
as close as 10~parsec will subtend angles $\ltorder 50$   
microarcseconds; any reflected or thermal radiation that 
they might emit thus will be confused by normal photometric 
techniques with that from the parent star.  Only  
exceedingly large and close companions of high albedo would be capable of 
creating a significant modulated signal throughout their orbit as the viewer 
sees a different fraction of the starlit side; we will not consider 
such planets here.  All other planets will alter the total observed 
flux only during an actual transit of the 
stellar face, during which the amplitude and shape of the photometric dip 
will be determined by the fraction of the stellar light that is occulted 
as a function of time.

\vglue -0.5cm
\hglue 2.5cm
\epsfxsize=8cm\epsffile{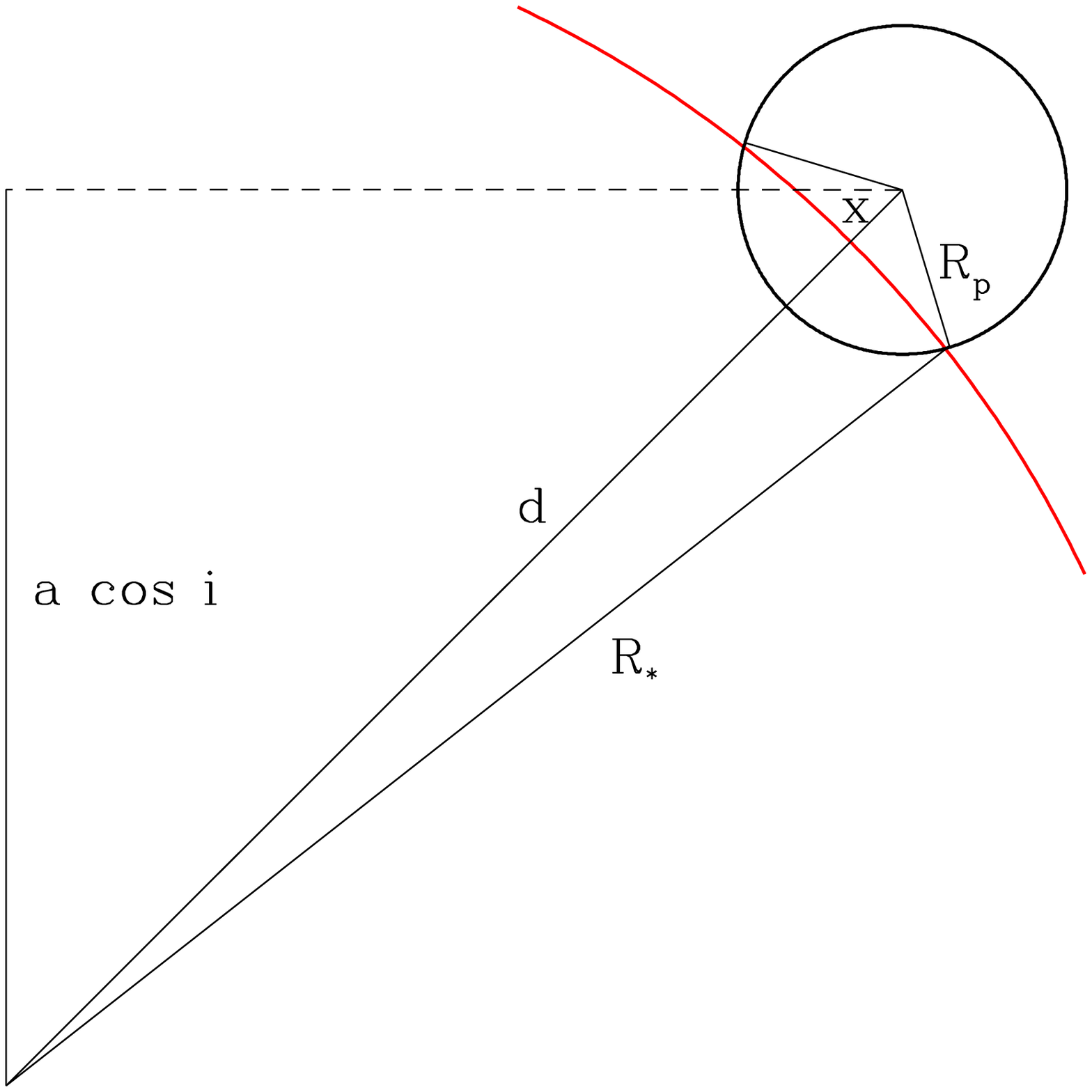}

\vglue -0.5cm{\small Fig.~7 --- The area eclipsed by a planet 
as it crosses the stellar limb determines the wing shape of the resulting 
photometric dip.}
\vskip 0.4cm

The maximum fractional change in the observed flux is given by:
\begin{equation}
{\rm Maximum~~} \frac{\abs \delta {\cal F_\lambda} \abs}{{\cal F_\lambda}} = 
\frac{\pi {\cal F}_{\lambda, *} \, R_p^2}
{\pi {\cal F}_{\lambda, *} \, R_*^2 + \pi {\cal F}_{\lambda, p} \, R_p^2} 
\approx \left(\frac{R_p}{R_*}\right)^2 \equiv \rho^2
\end{equation} 

\noindent
The shape of the transit dip will depend on the inclination angle, 
the ratio of the planet to stellar size, and the degree of limb-darkening 
in the observational band.  

Begin by considering a star of uniform 
brightness (no limb-darkening) transited by a small planet.
The stellar limb will then describe a nearly straight 
chord across the planet at any time, and integration over 
planet-centered axial coordinates (see Fig.~7) yields an eclipsing area  
during ingress and egress of:
\begin{equation}
{\cal A_E}    
\approx \int_x^{R_p} r_p \, dr_p \,
\int^{+ \arccos{(x/r_p)}}_{- \arccos{(x/r_p)}} 
\, d\phi_p \, \, = 2 \, \int^{R_p}_x r_p \, \arccos{\left(\frac{x}{r_p}\right)} \, dr_p ~~,
\end{equation} 

\noindent
where $x \equiv d - R_*$, $d$ is the projected star-planet separation 
and $x$ is constrained to lie in the region $ - R_p < x < R_p$.  
The last integral can be done analytically to yield,
\begin{equation}
{\cal A_E}   
\approx R_p^2 \, \arccos{(x/R_p)} - R_p x \, \sqrt{1 - \frac{x^2}{R_p^2}} ~~~.
\end{equation} 

For larger planets, and to facilitate the introduction of 
limb-darkened sources, it is more useful to 
integrate over stellar-centered axial coordinates; the Law of Cosines can 
then be used to show that 
\begin{equation}
{\cal A_E}(t)   
= 2 \int^{^{{\rm min}(R_*, \, d(t) + R_p)}}_{_{{\rm max}(0, \, d(t) - R_p)}} r_*  \, 
\arccos{[\Theta(t)]}  \, dr_* 
\end{equation} 
\begin{equation}
{\rm where ~~~}   
\Theta(t) \equiv  
\frac{d^2(t) + r_*^2 - R_p^2}{2 r_* d(t)} ~~~~ {\rm for~}r_* > R_p + d(t)
{\rm ,~and~}\pi{\rm ~otherwise.}
\end{equation} 

The light curve resulting from the occultation of a uniform brightness 
source by a planet of arbitrary size, orbital radius and orbital inclination 
can now be constructed by substituting into Eq.~9 the 
time dependence of the projected planet-star separation,   
$d(t) =$ $a \, \sqrt{\sin^2{\omega t} + \cos^2{i} \cos^2{\omega t}}$, 
where $\omega \equiv 2\pi/P$.  The {\it differential\/} light curve is then 
given by:
\begin{equation}
\frac{{\cal F} (t)}{{\cal F}_{0}} = 1 \, - \, 
\frac{{\cal A_E}(t)}{\pi \, R_*^2}
\end{equation} 

\noindent
For spherical stars and planets, the light curve will be symmetric 
and have a minimum at the closest projected approach of planet 
to star center, where the fractional decrease in 
the total brightness will be less than or equal to $(R_p/R_*)^2$.  
For Jupiter-sized planets orbiting solar-type stars, this is a 
signal of $\sim$1\%; 
for Earth-sized planets the fractional 
change is $\ltorder$ 0.01\% (Fig.~8).  

\vglue -4.3cm
\hglue 0.2cm
\epsfxsize=11cm\epsffile{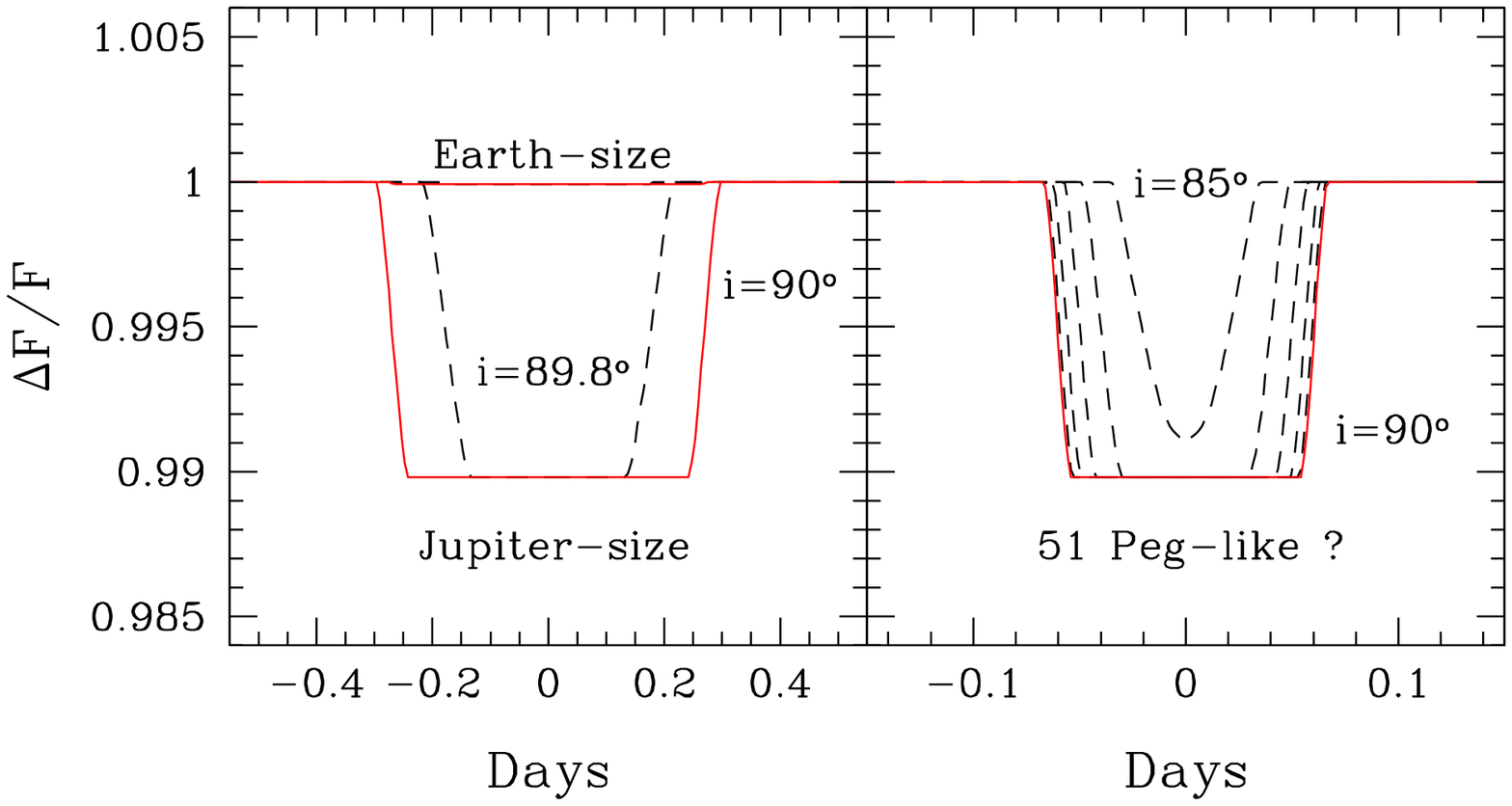}

\vglue -0.6cm{\small Fig.~8 --- {\bf Left:} Photometric light curves for 
Earth-sized and Jupiter-sized planets orbiting a solar-type star at 1 AU. 
{\bf Right:} A Jupiter-sized planet orbiting a solar-type star at an 
orbital radius of 0.05 AU (\eg\ 51 Peg) with  
inclinations ranging from 85\deg\ to 90\deg.  
The parent star is assumed here to have 
constant surface brightness.  Note change in time scale between two panels.}
\vskip 0.4cm

\noindent 
If proper care is taken, photometry of bright, uncrowded 
stars can be performed to $\sim$0.1\% precision from the ground 
(Henry \etal\ 1997), 
so that ground-based transit searches can in principle be sensitive 
to Jupiter-sized planets at $\ltorder$1 AU --- planets perhaps similar 
to those being found by the radial velocity technique 
(\eg\ Mayor \& Queloz 1995, Butler \& Marcy 1996). 
Transit detections of terrestrial planets like those in our own 
Solar System must await space observations in order to achieve the 
required photometric precision.

\subsubsection{Effects of Limb Darkening}

Because observations at different wavelengths probe material at different 
depths in stellar atmospheres, a stellar disk is differentially limb-darkened: 
the radial surface brightness profile $B_\lambda (r_*)$ of a star is 
wavelength dependent.   In redder bands, which probe the cooler 
outer regions of the star, the stellar disk will appear larger and 
less limb-darkened.  Limb darkening is important to transit techniques 
for two reasons:  it changes the shape of the photometric signal 
and it does so in a wavelength-dependent way.

Since a given planet can produce dips of varying strength depending on the 
inclination $i$ of the orbit, the inclination must be known in order to 
estimate the planet's radius $R_p$ accurately.
In principle, if the parent star has been typed so that its mass and stellar 
radius $R_*$ are known, Kepler's Law together with Eq.~5 will 
yield $i$ once the transit time $t_T$ and period $P$ have been measured.  
Ignoring the effects of limb darkening, however, 
will result in an underestimate of $t_T$, and thus an underestimate 
for the inclination $i$ as well.  
In order to produce the required amplitude at minimum, the size of 
the planet $R_p$ will then be overestimated.  
Furthermore, the sloping shape of the limb-darkened profile might be 
attributed to the smaller inclination $i$, reinforcing misinterpretation.   

This difficulty will be removed if the limb darkening can be properly 
modeled.  In addition, transit monitoring 
in more than one waveband could confirm the occultation hypothesis 
by measuring the characteristic color signature associated with 
limb darkening.  
In principle this signature can be used to determine the 
orbital inclination from a single transit, in which case Eq.~5 can be 
inverted to solve for the period $P$ without waiting for a second 
transit.

How strong is the effect of limb darkening? To incorporate its effect, 
the integral in Eq.~9 used to determine the eclipsing area must 
be weighted by the surface brightness as a function of stellar radius,   
yielding the differential light curve: 
\begin{equation}
\frac{{\cal F_\lambda} (t)}{{\cal F_\lambda}_{, \, 0}} = 1 \, - \, 
\frac{\int^{^{{\rm min}(R_*, \, d(t) + R_p)}}_{_{{\rm max}(0, \, d(t) - R_p)}} r_*  \, B_{\lambda}(r_*) \,
\arccos{[\Theta(t)]}  \, dr_*}{\pi \int^{^{R_*}}_{_0} r_*  \, B_{\lambda}(r_*) 
\, dr_*}
\end{equation} 

A commonly-used functional form for the surface brightness profile 
is $B_{\lambda}(\mu) = [1 - c_\lambda (1-\mu)]$, where 
$\mu \equiv \cos{\gamma}$ and $\gamma$ is the angle between the 
normal to the stellar surface and the line-of-sight.  In terms of the 
projected radius $r_*$ from the stellar center this can be written as  
$B_{\lambda}(r_*) = [1 - c_\lambda (1 - \sqrt{1 - (r_*/R_*)^2})]$.
Using this form and constants $c_\lambda$ appropriate for the Sun, 
light curves and color curves are shown in Fig.~9 for a Jupiter-sized 
planet orbiting 1~AU from a solar-type star at inclinations of 90\deg\ 
and 89.8\deg.  

As expected, the bluer band shows more limb darkening,  
which rounds the sharp edges of the occultation profile making it 
qualitatively degenerate with a larger planet at somewhat smaller 
inclination.  The color curves for different inclinations, however, 
are qualitatively different and can thus be used to break this 
degeneracy.  During ingress and egress the color curve becomes bluer 
as the differentially redder limb is occulted; at maximum occultation 
the color curve is redder than the unocculted star for transits 
with inclination $i = 90$\deg\ since the relative blue central 
regions are then occulted.  For smaller inclinations, the planet 
grazes the limb blocking preferentially red light only, and the 
color curve stays blue through the event.  Since the size of the 
color signal is $\sim$10\% of the deviation in total flux, excellent 
photometry is required to measure this effect and use it to estimate 
the orbital inclination; even for jovian giants it remains at 
or just beyond the current limits of photometric precision. 

\vglue -0.3cm
\hglue 0cm
\epsfxsize=11cm\epsffile{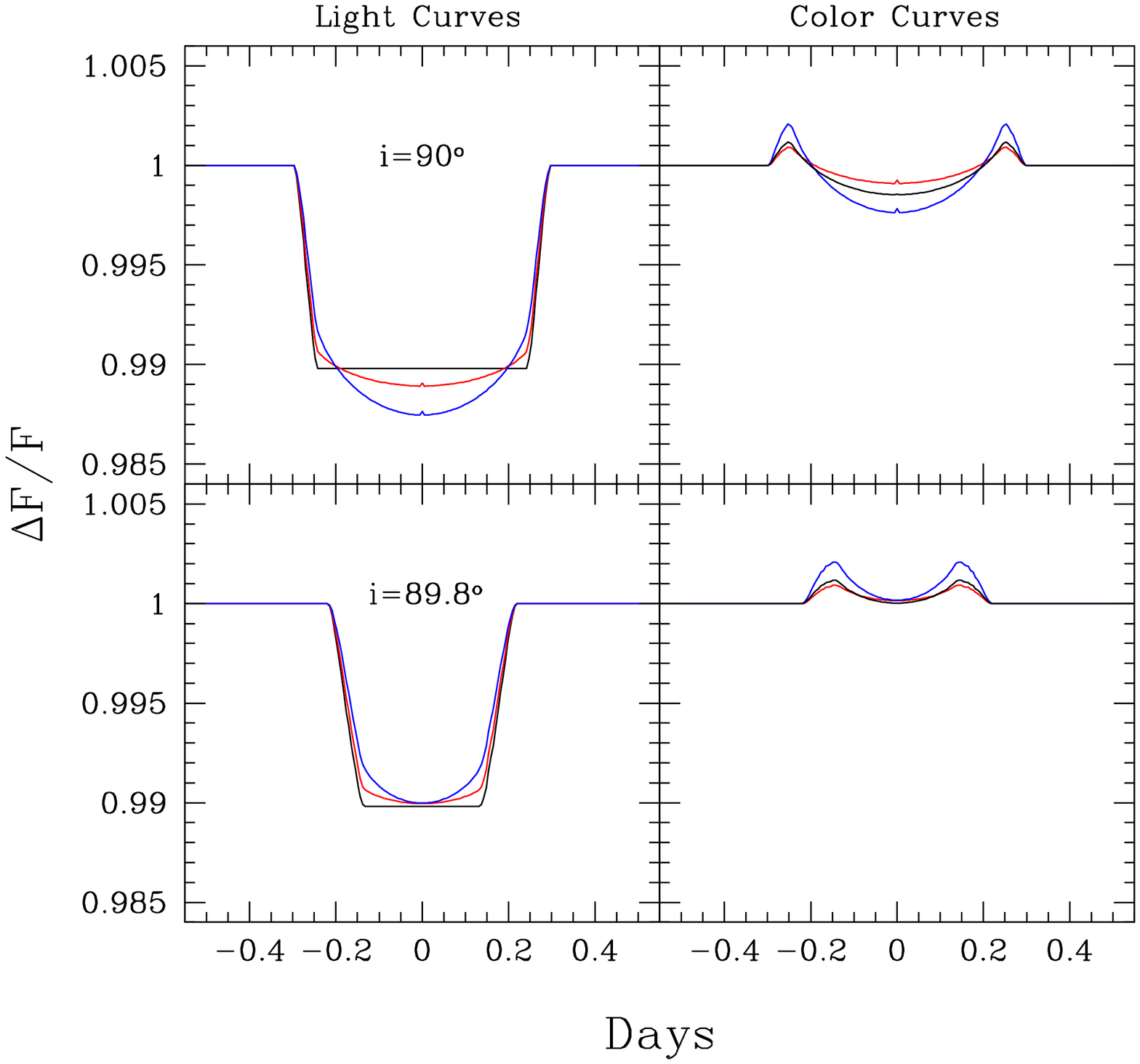}

\vglue -0.4cm{\small Fig.~9 --- {\bf Left:} Light curves for a 
planet with $R_p = 11 R_\earth$ orbiting a solar-type star with  
orbital inclinations of 90\deg\ (top) and 89.8\deg\ (bottom) normalized 
to the total (unocculted) flux in the indicated band. Black 
shows a uniformly bright stellar disk; blue and red indicate observations 
in the R and K bands respectively.  {\bf Right:} Color curves indicating 
the flux ratios at any given time between R (blue) and K-band (red) 
limb-darkened curves and a uniformly bright target star, and the 
observed limb-darkened $R/K$ flux ratio (black).}
\vskip 0.4cm

\subsection{Observational Rewards and Challenges}

In sum, what can be learned by observing a planetary object transiting 
the face of its parent star?  The amplitude of the photometric signal 
places a lower limit on the ratio of the planetary radius to stellar 
radius $\rho \equiv R_p/R_*$, while the duration of the event 
places a lower limit on the orbital period $P$ and thus on 
the orbital radius $a$ as well.  If the inclination $i$ is known, 
these lower limits become measurements.  In principle $i$ could 
be determined by fitting the wings of the transit profile 
in different wavebands using the known limb-darkening of the star, 
but in practice this will probably prove too difficult.  Instead, 
multiple transits will be required to time the transits and 
thus measure the period $P$ of the planet, from which the inclination can 
be determined from the known transit duration (Eq.~5).  This makes the 
transit method most appropriate for large planets orbiting their 
parent stars at (relatively) small radii $a$.  The primary 
challenge then reduces to performing the very precise photometry  
required on a large enough sample of stars to place meaningful 
statistics on the numbers of planets at small $a$. 

What limits the photometric accuracy and clear detection of 
a transit signal?   The dwarf stars that have suitably small 
stellar radii $R_*$ must have apparent magnitudes bright enough 
(ie, be close enough) that enough photons can be captured in 
short exposures so that a sub-day transit event can be well-sampled. 
This will limit the depth of the sample to only nearby stars.  
Fields with very high stellar densities 
(like globular clusters or the Galactic Center) or very wide fields 
that can capture hundreds of candidate stars simultaneously will  
be required in order to maintain the required temporal sampling on a large 
enough sample.  Regions of high stellar density, however, will be hampered by 
the additional challenges associated with precision photometry in 
confused fields.  

The use of reference constant stars in the field can reduce the effects 
of varying extinction to produce the best current photometry in uncrowded 
fields, precise to the $\sim$0.1\% level.  
Ultimately, scintillation, the rapidly-varying turbulent refocusing of rays 
passing through the atmosphere, limits Earth-bound photometry 
to 0.01\%.  Detection of Earth-mass transits is thus probably 
restricted to space-borne missions, although in special circumstances, 
periodicity analyses may be used to search for very short-period 
Earth-sized transits from the ground (\eg Henry \etal\ 1997). 

For larger, jovian gas-giants, the signal can be measured from 
the ground, but must be distinguished from intrinsic effects that 
could be confused with transits.  Late-type dwarf stars often undergo 
pulsations that cause 
their brightness to vary on the order of a few hours, but 
due to their cyclic nature these pulsations should 
be distinguished easily from all but very short period transits 
corresponding to $a \ltorder 0.02$~AU or so.  

Solar flares produce excess of flux at the 
$\ltorder 0.001$\% level, and thus would not confuse a typical transit signal. 
Later-type dwarfs tend to have more surface activity, however, and 
thus produce flares that contain a larger fraction of the star's 
total flux.  Since the flares are generally blue, 
the primary problem will be in confusing the chromatic 
signal expected due to limb-darkening effects during a transit. 

More troublesome will be 
separating transits from irregular stellar variability due to star spots.  
Star spots are cool regions on the stellar 
surface that remain for a few rotations 
before disappearing.  They could mock a transit event and thus are 
probably the most important non-instrumental source of noise.  
Although the power spectrum of the Solar flux does show variations on 
day and sub-day time scales, most of the power during periods of 
sunspot maximum occurs at the approximate 1-month time scale of 
the Sun's rotation.  Even during sunspot maximum, variations on 
day and sub-day scales are at or below the $0.001\%$ level 
(Borucki, Scargle \& Hudson 1985).  
Star spots on solar-type stars will therefore not be confused with 
the transit signal of a gas giant, but spots might be a source of 
additional noise for terrestrial-sized planets of small 
orbital radius ($a \ltorder 0.3$AU) and for parent stars that are 
significantly more spotted than the Sun.
 
\subsubsection{Pushing the Limits: Rings, Moons and Multiple Planets}

If the parent star can be well-characterized, the transit method 
involves quite simple physical principles that can perhaps be 
exploited further to learn more about planetary systems.   
For example, if a system is discovered to contain large transiting 
inner planets, it can be assumed to have a favorable inclination 
angle that would make it a good target for more sensitive searches for 
smaller radius or larger $a$ planets in the same system.  

If the inner giants are large enough, differential spectroscopy 
with a very large telescope before and during transits could reveal 
additional spectral lines that could be attributed to absorption of 
stellar light by the atmosphere of the giant (presumably  
gaseous) planet (see Laurent \& Schneider, this proceedings).  
A large occulting ring inclined to the observer's 
line-of-sight would create a transit profile of a different shape 
than that of a planet (Schneider 1997), 
though the signal could be confused with 
limb-darkening effects and would likely be important only for 
outer gas giants where icy rings can form more easily.

Finally, variations in the ingress timing of inner planets 
can be used to search for cyclic 
variations that could betray the presence of moons (Schneider 1997) 
or --- in principle ---
massive (inner or outer) planets that are nearly coplanar but 
too misaligned to cause a detectable 
transit themselves.  Transit timing shifts would be caused  
by the slight orbital motion of the planet around the planet-moon 
barycenter or that of the star around the system barycenter.  
(The latter is unobservable for a single-planet system 
since the star's motion is always phase-locked with the planet.)

\newpage


\section{Principles of Planet Detection via Microlensing}

Microlensing occurs when a foreground compact object (\eg\ a star)  
moves between an observer and a luminous background object 
(\eg\ another star).  The gravitational field of the foreground lens 
alters the path of the light from the background source, bending 
it more severely the closer the ray passes to the lens.  This results  
in an image as seen by a distant observer that is altered both in 
position and shape from that of the unlensed source.  Indeed since light 
from either side of a lens can now be bent to reach the observer, 
multiple images are possible (Fig.~10).  Since the total flux reaching 
the observer from these two images is larger than that from the 
unlensed source alone, the lens (and any planets that may encircle it) 
betrays its presence not through its own luminous emission, but 
by its gravitational magnification of the flux of background objects.  
Einstein (1936) recognized microlensing in principle, but thought 
that it was undetectable in practice. 

\vglue -0.5cm
\hglue -1.5cm
\epsfxsize=9.75cm\epsffile{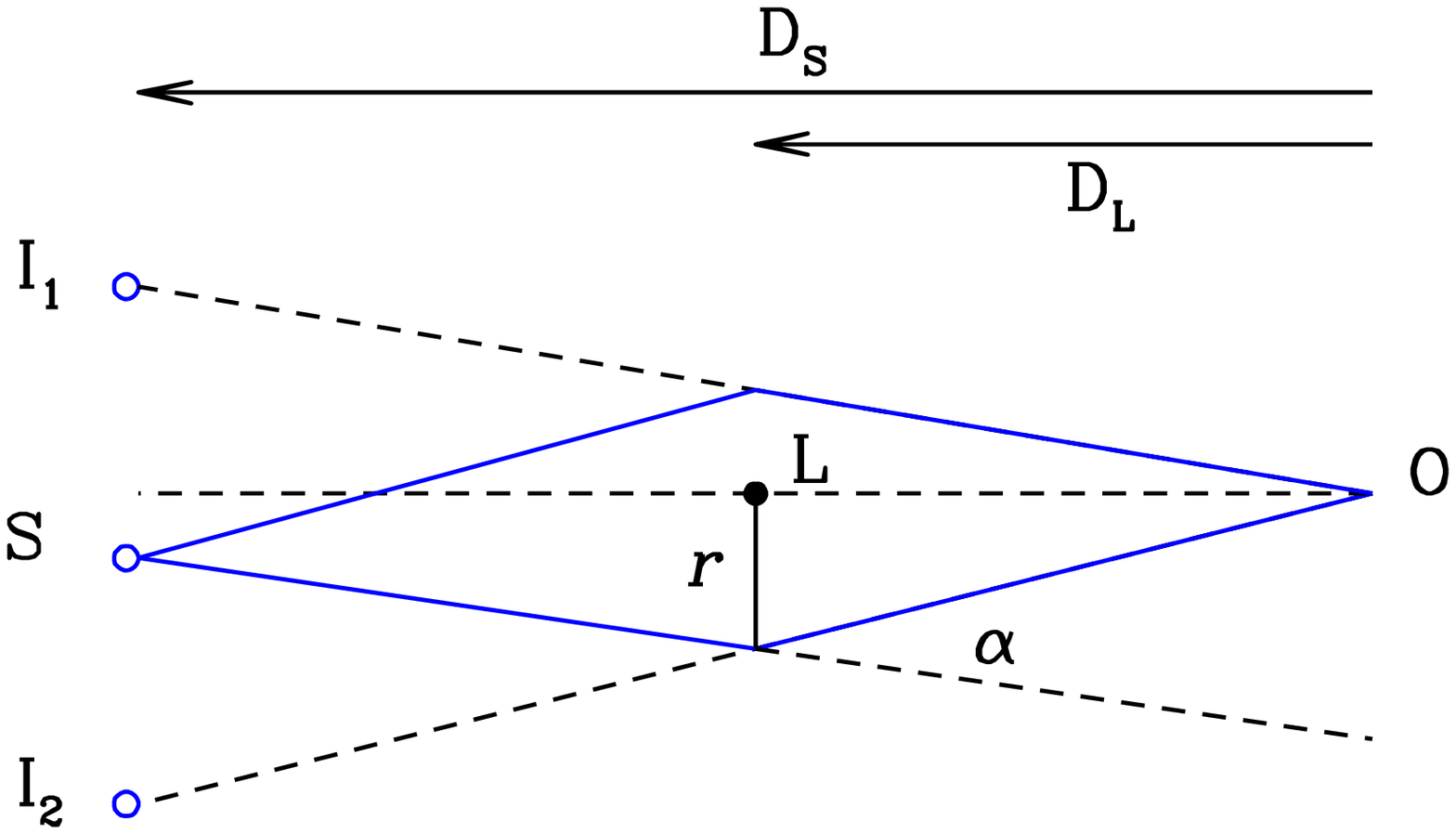}

\vglue -10.5cm
\hglue 4.75cm
\epsfxsize=9cm\epsffile{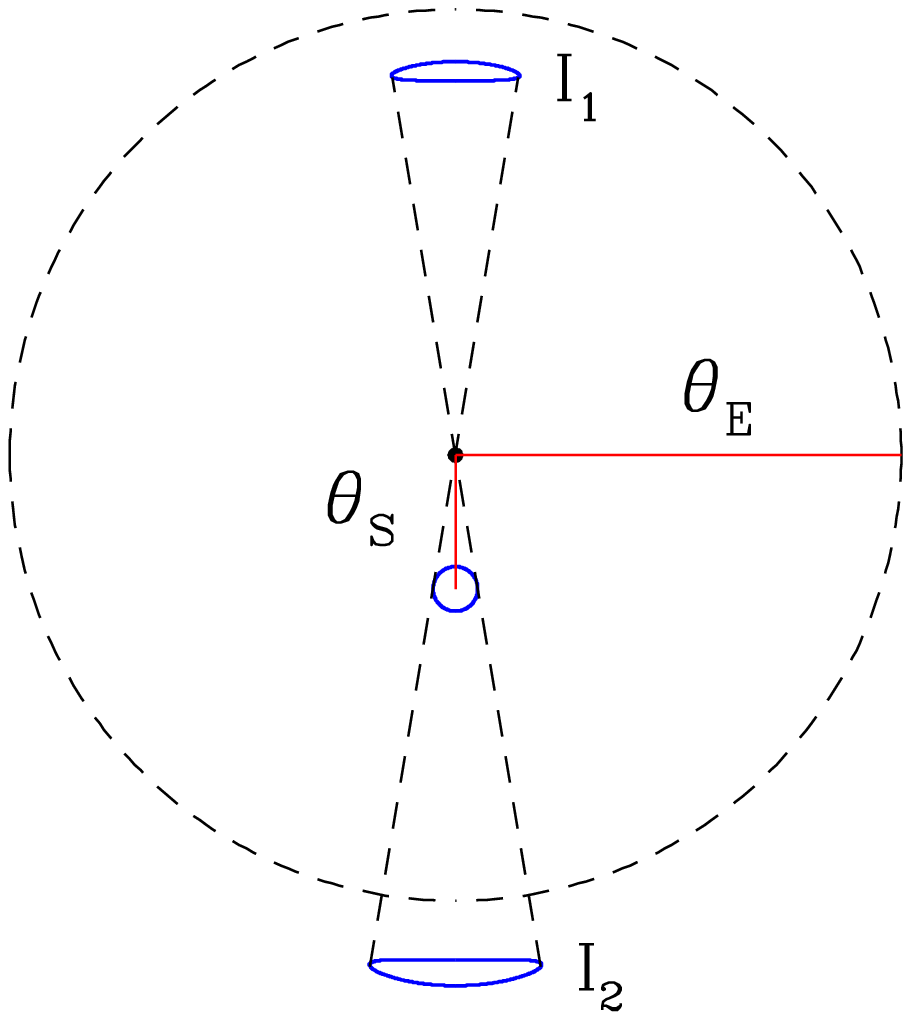}

\vglue -2.5cm{\small Fig.~10 --- {\bf Left:} A compact lens (L)
located a distance $D_L$ nearly along the line-of-sight to a background  
source (S) located at a distance $D_S$ will bend incoming light rays 
by differing amounts $\alpha$ to create two images 
($I_1$ and $I_2$) on either side of the line-of-sight. 
{\bf Right:} An observer $O$ does not see the microlensed 
source at its true angular sky position $\theta_S$, 
but rather two images at positions $\theta_1$ and $\theta_2$.}
\vskip 0.4cm

Ray tracing, together with the use of general relatively to relate 
the bending angle $\alpha$ with the lens mass distribution,   
produces a mapping from the source positions ($\xi$, $\eta$) to the 
image positions (x,y) for a given mass distribution. 
For ``point'' masses, the angle $\alpha$ is just given by the mass of the lens 
$M$ and the distance of closest approach $r$ as:
\begin{equation}
\alpha = \frac{4 \, G \, M}{c^2 \, r} = \frac{2R_S}{r}~~, 
\end{equation} 
as long as $r$ is well outside the Schwarzschild radius $R_S$ of the lens.
Simple geometry alone then requires
\begin{equation}
{\theta_S} \, D_S = { r} \, \frac{D_S}{D_L} - (D_S - D_L) \, {\alpha(r)} ~~, 
\end{equation} 
which can be rewritten to yield the lens equation
\begin{equation}
{\theta_S} = {\theta} - \frac{D_{LS}}{D_S} \, {\alpha(r)} ~~, 
\end{equation} 
giving the (angular) vector image positions $\bf \theta$ for a source 
at the angular position $\theta_S$ as measured from the observer-lens 
line-of-sight.  $D_S$ and $D_L$ are the source and lens distances 
from the observer, respectively, and $D_{LS} \equiv D_S - D_L$.

For convenience, the characteristic angular size scale is defined as
\begin{equation}
{\theta_E} \equiv \sqrt{\frac{2 R_S D_{LS}}{D_L \, D_S}} 
= \sqrt{\frac{4 G M D_{LS}}{c^2 \, D_L \, D_S}} ~~~. 
\end{equation} 
Since $r = D_L \, \theta$, Eq.~15 can now be rewritten to yield 
a quadratic equation in $\theta$
\begin{equation}
\theta^2 - \theta_S \, \theta \ - \theta_E^2 = 0 ~~~, 
\end{equation} 
with two solutions $\theta_{1, \, 2} = 
\frac{1}{2} \left( \theta_S \pm \sqrt{4 \theta_E^2 + \theta_S^2} \, \right) $ 
giving the positions of images $I_1$ and $I_2$.
When the source lies directly behind the lens as seen from the observer, 
$\theta_S = 0$ and the two images merge into a ring of radius $\theta_E$, 
the so-called ``Einstein ring.''  For all other source positions, one image 
will lie inside $\theta_E$ and one outside.  The flux observed from 
each image is the integral of the image surface brightness 
over the solid angle subtended by the (distorted) image.  Since the specific 
intensity of each ray is unchanged in the bending process, so is the 
surface brightness.  The magnification $A_{1, 2}$ for each image 
is then just the ratio of the image area to the source area,  
and is found formally by evaluating at the image positions the  
determinant of the Jacobian mapping $J$ that describes the lensing 
coordinate transformation from image to source plane:
\begin{equation}
A_{1, \, 2} = \left. \frac{1}{\abs det \, J \abs} \right|_{\, \theta = \theta_{1, \, 2}} 
= \left| \frac{\partial \, {\theta_S}}{\partial \, {\theta}} \right|^{-1}_{\, \theta = \theta_{1, \, 2}}~~~, 
\end{equation} 
where $\theta_S$ and $\theta$ are (angular) position vectors for the 
source and image, respectively.

What is most important for detection of extra-solar planets around lenses 
is not the position of the images but their magnification.   
For stellar lenses and typical source and lens distances 
within the Milky Way, the typical image separation ($\gtorder 2\theta_E$) 
is $\sim$1~milliarcsecond, too small 
to be resolved with current optical telescopes.   The observer 
sees one image with a combined 
magnification $A \equiv A_1 + A_2$ that can be quite large.   
In order to distinguish intrinsically 
bright background sources from fainter ones that appear bright due to 
microlensing, the observer relies on the characteristic 
brightening and dimming that occurs as motions within the Galaxy  
sweep the source (nearly) behind the lens-observer line-of-sight.  
The unresolved images also sweep across the sky (Fig.~11); their combined 
brightness reaches its maximum when the source has its closest 
projected distance to the lens. 

\vglue -1.75cm
\hglue 1cm
\epsfxsize=10cm\epsffile{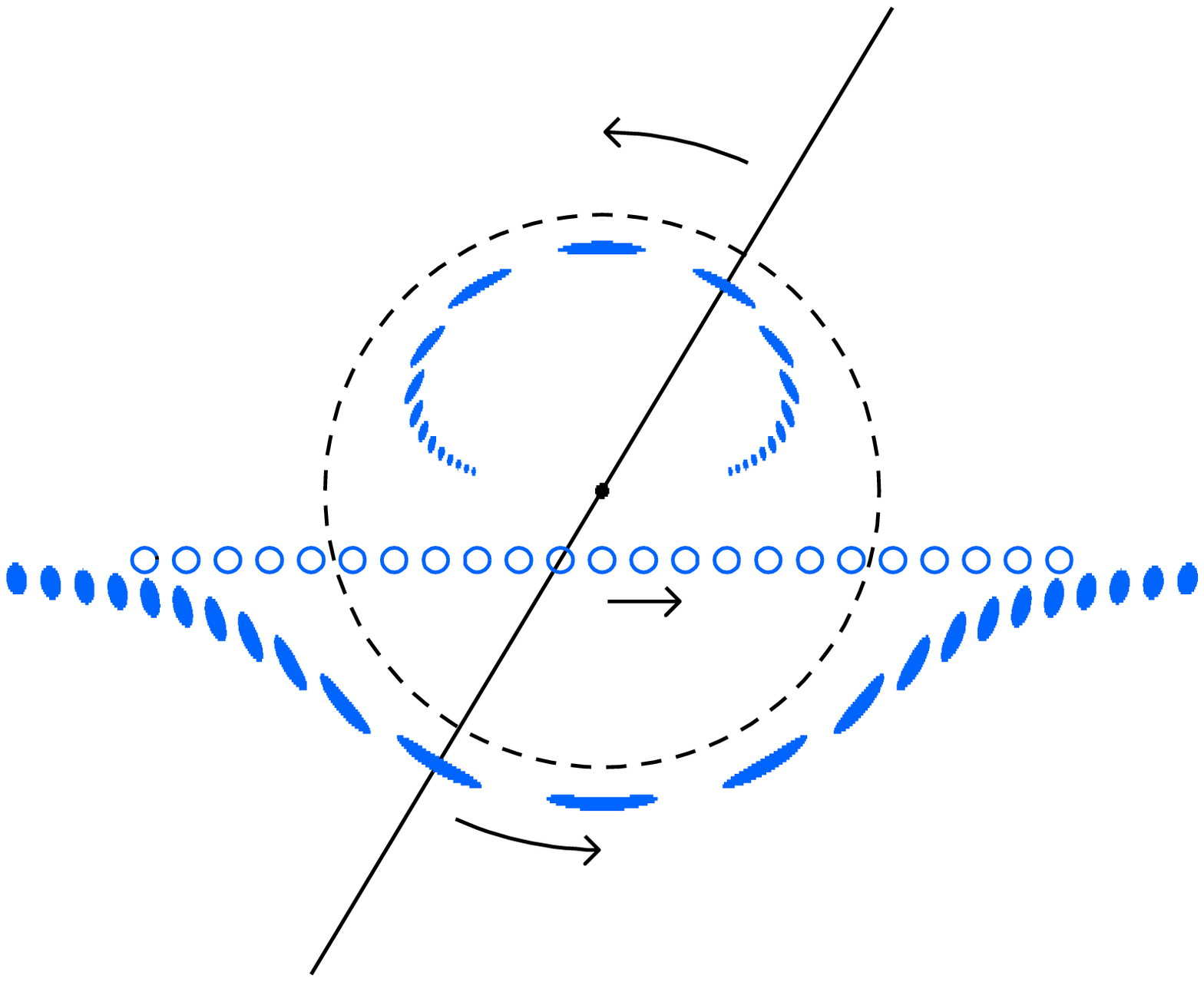}

\vglue -1.75cm{\small Fig.~11 --- As a background source 
(open circle) moves nearly behind a foreground lens (central dot), 
the two microimages remain at every moment colinear with the lens 
and source.  (Adapted from Paczy\'nski 1996.)}
\vskip 0.4cm

For a single lens, the combined magnification 
can be shown from Eqs.~17 and 18 to be:
\begin{equation}
A = \frac{u^2+2}{u \sqrt{u^2+4}} ~~~,
\end{equation}
where $u \equiv \theta_S/\theta_E$ is the angular source-lens separation 
in units of the Einstein ring radius.  For rectilinear motion, 
$u(t) = \sqrt{(t - t_0)^2/t_E^2 + u^2_{min}}$, where 
$t_0$ is the time at which $u$ is minimum and the magnification is maximum, 
and $t_E \equiv \theta_E \, D_L / v_\perp$ is the characteristic 
time scale defined as the time required for the lens to travel a 
projected distance across the observer-source sightline 
equal to the Einstein radius $r_E$.
The result is a symmetric light curve that has a magnification 
of 1.34 as it cross the Einstein ring radius and a peak amplification 
that is approximately inversely proportional to the source impact 
parameter $u_{min}$.  Since the $u_{min}$ are distributed randomly, 
all of the light curves shown in Fig.~12 are equally probable.

\vglue -0.3cm
\hglue -2.25cm
\epsfxsize=10.5cm\epsffile{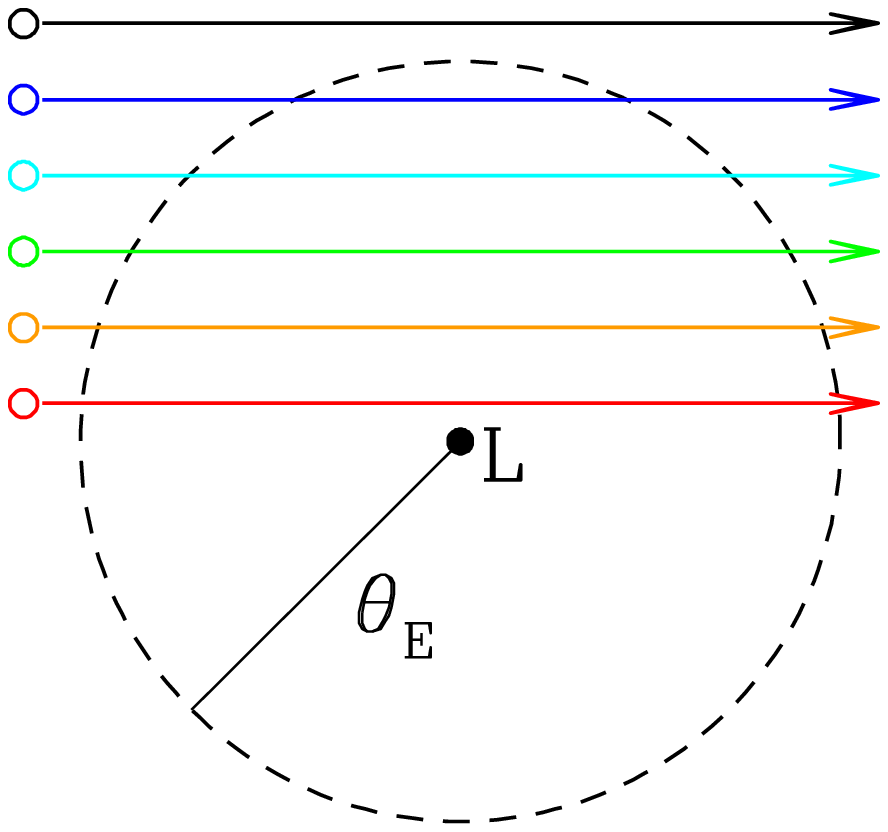}

\vglue -10cm
\hglue 4.5cm
\epsfxsize=8cm\epsffile{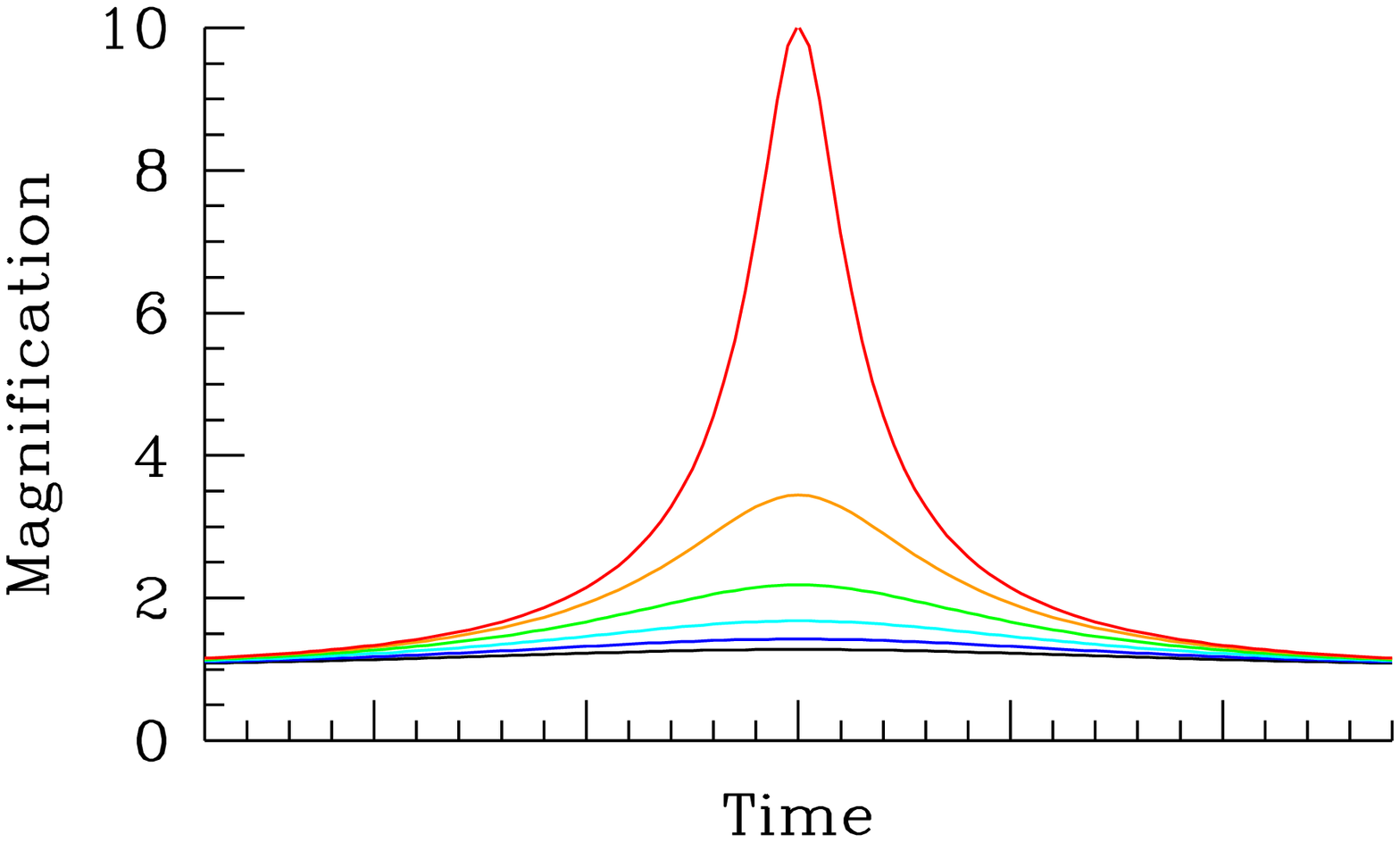}

\vglue -3cm{\small Fig.~12 --- {\bf Left:} Equally-probable source 
trajectories.  {\bf Right:} The corresponding single microlens light curves.}
\vskip 0.4cm

Typical event durations $\hat t = 2 t_E$ 
for microlensing events detected in the direction of 
the Galactic Bulge are on the order of a few weeks to a few months, 
generally matching expectations for stellar lenses distributed in 
the Galactic disk and bulge.

\subsection{Microlensing by Binary Lenses}

Microlensing was proposed as a method to detect compact baryonic 
dark matter in the Milky Way by Paczy\'nski in 1986.  In 1991, 
Mao and Paczy\'nski suggested that not only dark lenses, but possible 
dark planets orbiting them may be detected through their microlensing 
influence on background stars.

The magnification patterns of a single lens are axially symmetric 
and centered on the lens; the Einstein ring radius, for example, 
describes the position of the $A \equiv A_1 + A_2 = 1.34$ 
magnification contour. 
Binary lens structure destroys this symmetry:  
the magnification patterns become distorted and are symmetric only 
upon reflection about the binary axis.    
Positions in the source place for which the 
determinant of the Jacobian (Eq.~18) is zero represent potential 
source positions for which the magnification is formally infinite.
The locus of these positions is called a ``caustic.''  For a 
single point-lens, the only such position is the point caustic 
at $\theta_S = 0$, but the caustics of binary lenses are extended 
and complicated in shape.  In the lens plane, the condition $|{\rm det}~J| = 0$ 
defines a locus of points known as the critical curve; when the source 
crosses a caustic a pair of new images of high amplification appear 
with image positions ${\bf \theta}$ on the critical curve.

A static lens configuration has a fixed magnification pattern 
relative to the lens; 
the observed light curve is one-dimensional cut through this pattern that  
depends on the source path.  
As Fig.~13 illustrates, 
the exact path of the source trajectory behind a binary lens 
will determine how much its light curve deviates from the simple 
symmetric form characterizing a single lens.  Due to the finite size of the 
source, the magnification during a caustic crossing is not infinite, 
but will be quite large for sources that are small compared to the 
size of the caustic structure.  Several binary-lens light curves have 
already been observed and characterized (Udalski \etal\ 1994, Alard, Mao \& Guibert 1995, Alcock \etal\ 1997, Albrow \etal\ 1998b, Albrow \etal\ 1999).

\vglue -1cm
\hglue -1.25cm
\epsfxsize=8cm\epsffile{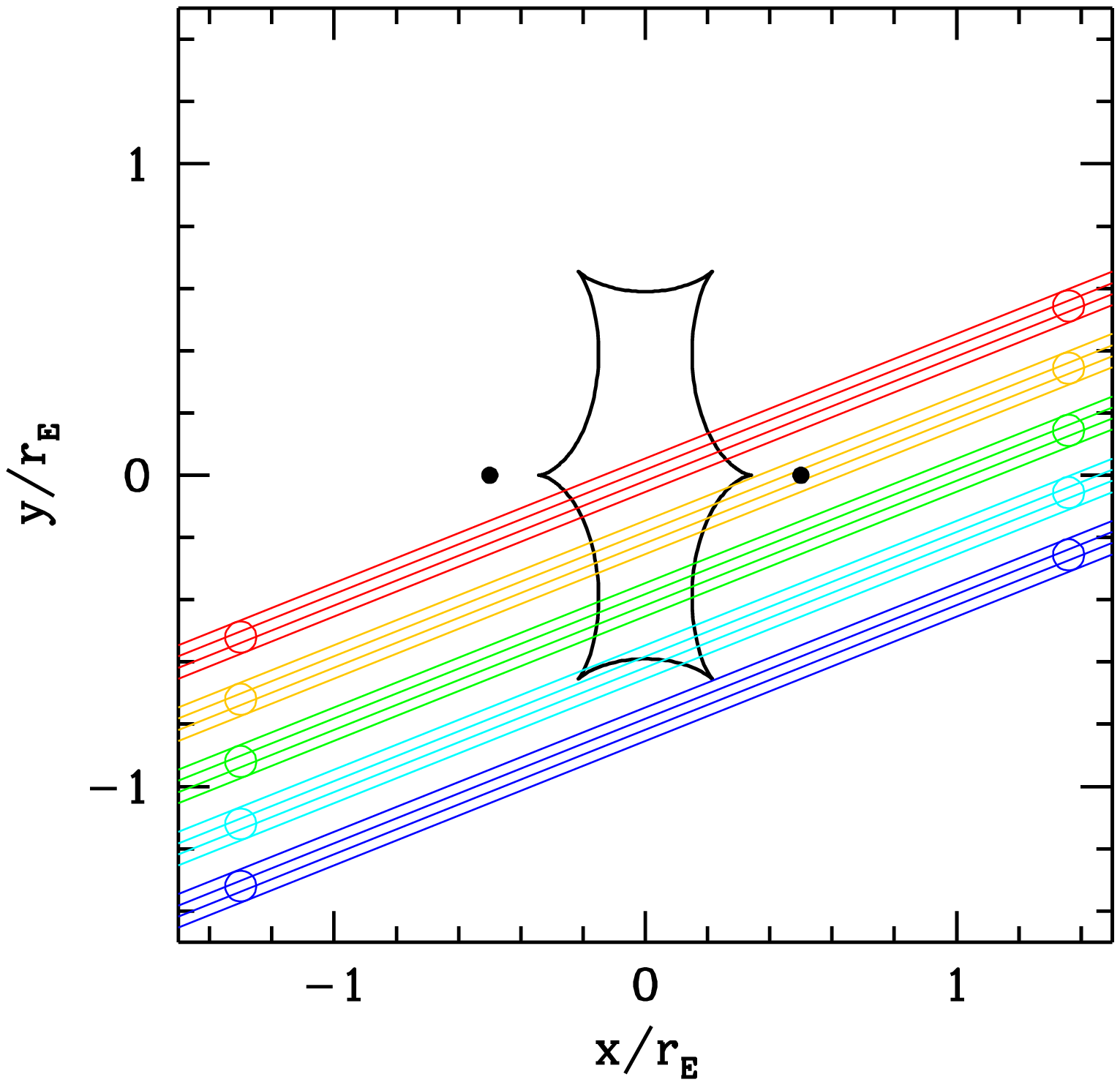}

\vglue -8.25cm
\hglue 5.25cm
\epsfxsize=8cm\epsffile{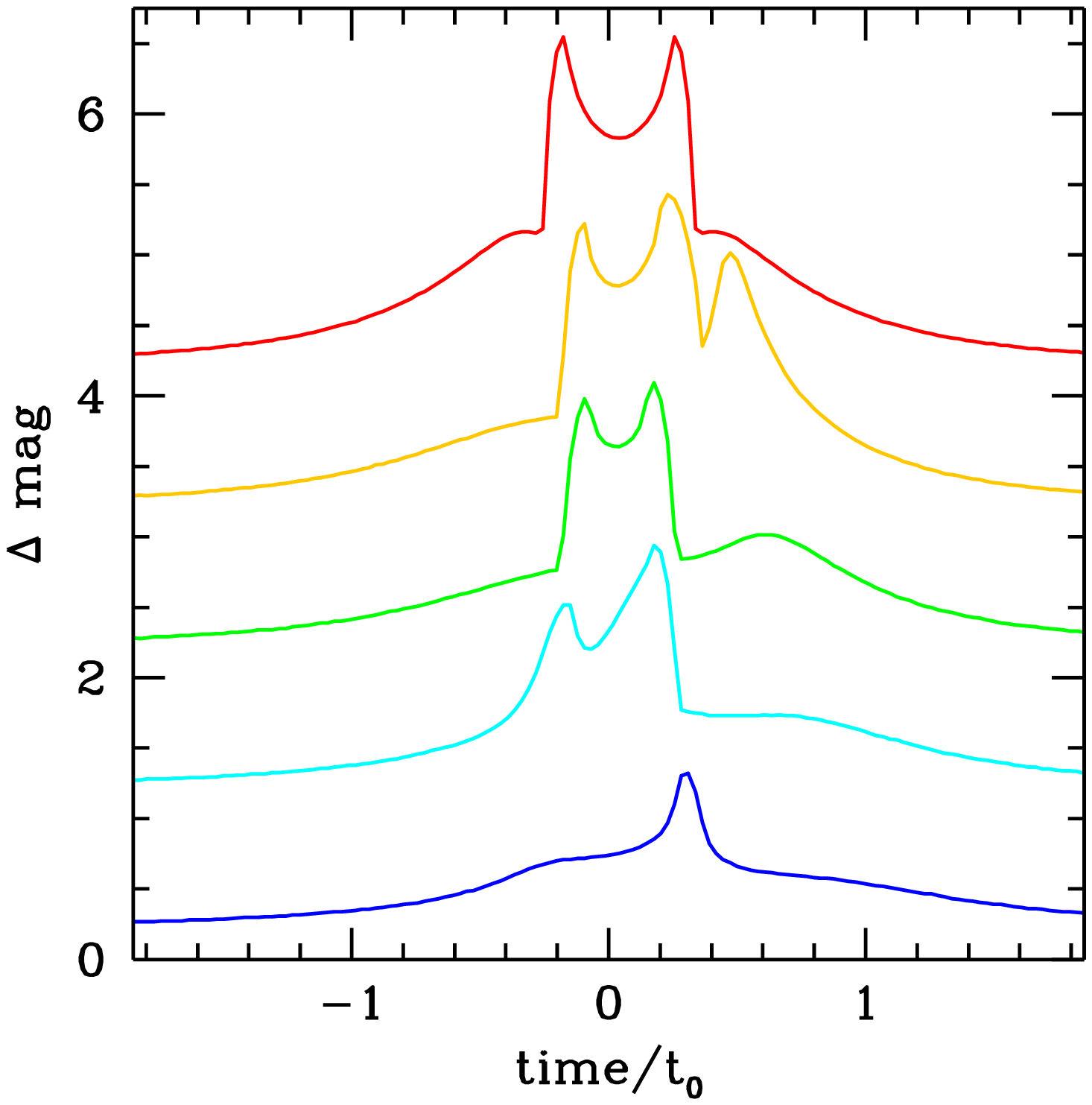}

\vglue -0.5cm{\small Fig.~13 --- {\bf Left:}  The caustic (thick closed line) 
for two equal mass lenses (dots) is shown with several possible source 
trajectories.  Angular distances are scaled to the Einstein ring radius of 
the combined lens mass.  {\bf Right:}  The light curves 
resulting from the source trajectories shown at left; the temporal axis 
is normalized to the Einstein time $t_E$ for the combined lens. (Adapted from 
Paczy\'nski 1996.)}
\vskip 0.4cm

A single lens light curve is described by four parameters: 
the Einstein crossing time $t_E$, the impact parameter $u_{min}$, 
the time of maximum amplification $t_0$, and 
the unlensed flux of the source $F_0$.   Only the first of these 
contains information about the lens itself.  Three additional 
parameters are introduced for binary lenses: 
the projected separation $b$ of the lenses in  
units of $\theta_E$, the mass ratio $q$ of the two lenses, and  
the angle $\phi$ that the source trajectory makes with respect 
to the binary axis.  
Given the large number of free parameters and the variety of complicated 
forms that binary light curves can exhibit, it may seem quite 
difficult to characterize the binary lens with any degree of
certainty on the basis of a single 1-D cut through its magnification 
pattern.  In fact, with good data the fitting procedure 
is unique enough that the {\it future\/} behavior of 
the complicated light curve --- including the timing of future 
caustic crossings --- can be predicted in real time.  This is important since 
the ability to characterize extra-solar planets via microlensing 
requires proper determination of the planetary system parameters 
$b$ and $q$ through modeling of light curve anomalies.

\subsection{Planetary Microlensing}

The simplest planetary system is a binary consisting of a 
stellar lens of mass $M_*$ orbited by a planet of mass $m_p$ at 
an orbital separation $a$.   
The parameter range of interest 
is therefore $q \equiv M_*/m_p \approx 10^{-3}$ for jovian-mass planets and 
$q \approx 10^{-5}$ for terrestrial-mass planets. 
The normalized projected angular separation $b \ltorder a/(\theta_E D_L)$ 
depends at any moment on the inclination and phase of the planetary orbit.
The light curve of a source passing behind a lensing planetary system 
will depend on the form of the magnification pattern of the lensing 
system, which is influenced by the size and position of the 
caustics.  How do the magnification patterns vary with $b$ and $q$? 

\hglue -1.5cm
\epsfxsize=14cm\epsffile{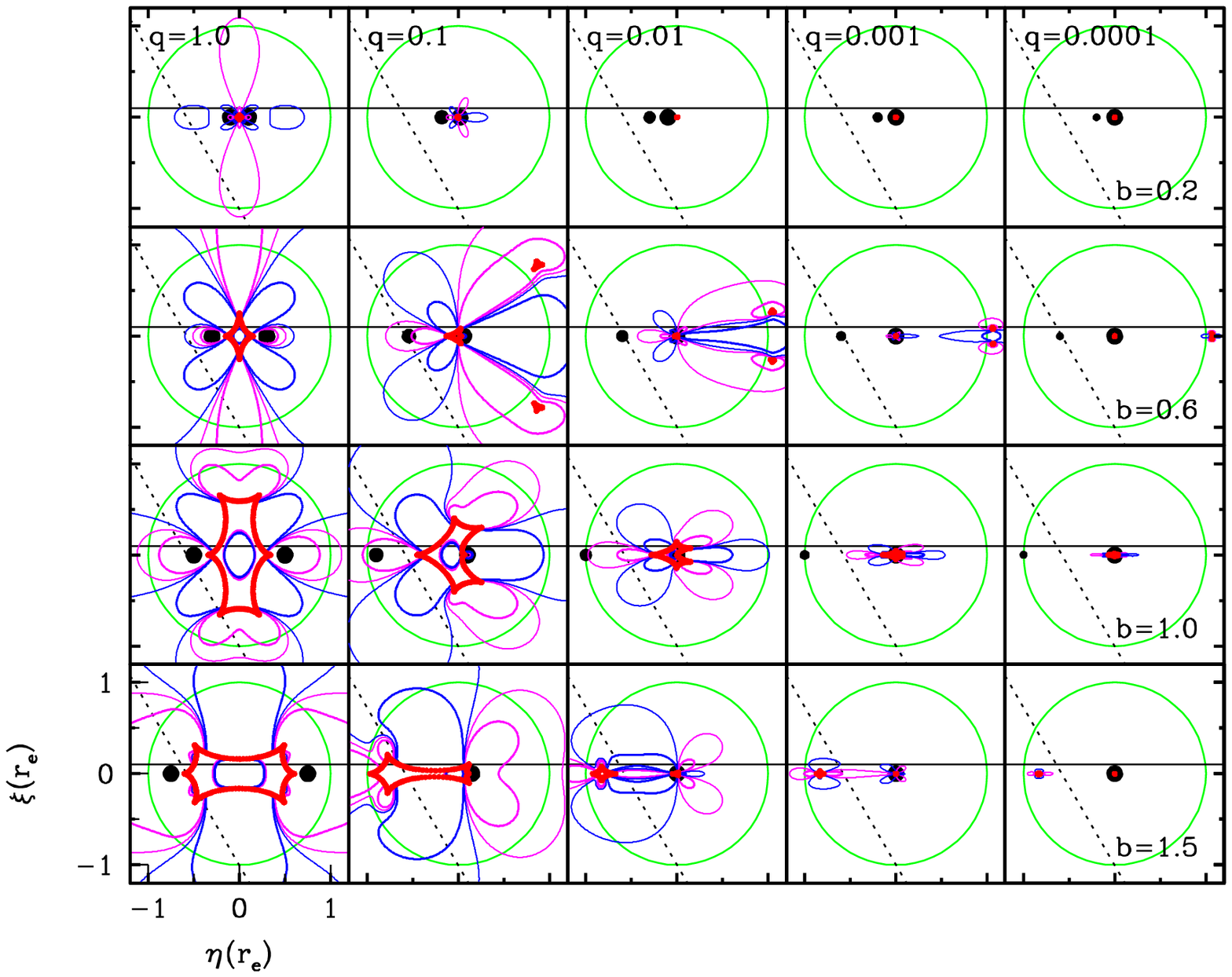}
\vglue -4.25cm{\small Fig.~14 --- 
Positive (magenta) and negative (blue) 1\% and 5\% {\it excess\/} 
magnification contours for binary lenses (black dots) 
of different projected separations $b$ and mass ratios $q$. 
Caustics are shown in red.  Dimensions are normalized 
to the Einstein ring radius of combined system (green circle).  
Dashed and solid lines are two possible source 
trajectories.  (Adapted from Gaudi \& Sackett 1998.)}
\vskip 0.4cm

Shown in Fig.~14 is the {\it excess\/} magnification pattern of a binary  
over that of a single lens for different separations $b$ 
and mass ratios $q$. 
The deviations can be positive or negative.  High-mass ratio binaries 
(\ie\ $q$ not too much less than 1) 
are easier to detect since their excess magnification contours 
cover a larger sky area 
making it more likely that a source trajectory will cross an 
``anomalous'' region.  For a given mass ratio $q$, the 1\% and 5\% 
excess magnification contours also cover more sky when the binary 
separation is comparable to the Einstein ring radius of the system, 
\ie\ whenever $b \approx 1$.  

The symmetric caustic structure centered between equal mass ($q = 1$) 
binaries becomes elongated along the binary axis for smaller mass 
ratios, eventually splitting the caustic into a central caustic 
centered on the primary lens and outer ``planetary'' caustics.  
For planetary separations larger than the Einstein ring radius $b > 1$, 
the planetary caustic is situated on the binary axis between the lens 
and planet.  For $b < 1$, the 
planetary caustics are two ``tips'' that are symmetrically positioned 
above and below the binary axis on the opposite side of the lens 
from the planet.  As the mass ratio decreases, all the caustics 
shrink in size and the two ``tips'' approach the binary axis, nearly ---
but not quite --- merging.

\subsubsection{The ``Lensing Zone''}

For the planetary (small $q$) regime, a source that 
crosses the central caustic will generate new images near the Einstein 
ring of the primary lens; a source crossing a planetary caustic will 
generate new images near the Einstein ring of the planet, \ie\ near 
the position of the planet itself.   Planets with separations 
$0.6 \ltorder b \ltorder 1.6$ create planetary caustics inside 
the Einstein ring radius of the parent lensing star; this is  
the region in which the source must be in order to be 
alerted by the microlensing survey teams.  
For this reason, planets with projected separations 
$0.6 \ltorder b \ltorder 1.6$ are said to lie in the 
{\it ``lensing zone.''\/} 
Since the separation $b$ is normalized to the size of the Einstein 
ring, the physical size of the lensing zone will depend on 
the lens mass and on the lens and source distances.  Most of 
the microlensing events in the Milky Way are detected in the direction 
of the Galactic bulge where, at least for the bright red clump sources, 
it is reasonable to assume that the sources lie at $D_S \approx 8 \, $kpc. 
Table I shows the size of the lensing zone for foreground 
lenses located in the disk ($D_L = 4 \, $kpc) and bulge ($D_L = 6 \, $kpc) 
for typical stellar masses, assuming that $D_S = 8 \, $kpc.

One of the reasons that microlensing is such an attractive method 
to search for extra-solar planets is that the typical lensing zone 
corresponds to projected separations 
of a few times the Earth-Sun distance (AU) --- a good match to many planets 
in the Solar System.   Planets orbiting at a radius $a$ 
in a plane inclined by $i$ with respect to the plane of the sky will 
traverse a range of projected separations 
$a \, \cos{i}/(\theta_E \, D_L) < b < a/ (\theta_E \, D_L)$, 
and can thus be brought into the lensing zone of their primary even 
if their orbital radius is larger than the values given in Table I. 

{\small 
\begin{center}
\begin{tabular}{l r r}
\noalign{\medskip\hrule\smallskip}
\multicolumn{3}{c}{TABLE I.  Typical Lensing Zones for Galactic Lenses}\\
\noalign{\hrule}
\noalign{\smallskip\hrule\smallskip}
\medskip
~~~~  		           &   disk lens (4 kpc)    &  bulge lens (6 kpc) \\

1.0 $\msolar$ solar-type   &      2.4 - 6.4 AU      &    2.1 - 5.5 AU \\

0.3 $\msolar$ dwarf        &      1.3 - 3.5 AU      &    1.1 - 3.0 AU \\
\noalign{\medskip\hrule\smallskip}
\end{tabular}
\end{center}
}

\vskip 0.25cm
Planets that are seldom or never brought into the lensing zone of 
their primary can still be detected by microlensing in one of two ways.
Either the light curve must be monitored for source positions outside 
the Einstein ring radius of the primary (\ie\ for magnifications $A < 1.34$) 
in order to have sensitivity to the isolated, outer planetary caustics  
(DiStefano \& Scalzo 1999), 
or very high amplification events must be monitored in 
order to sense the deviations that are caused any planet on the 
central primary caustic (Griest \& Safizadeh 1998).

\subsubsection{Determining the Planet-Star Mass Ratio and Projected Separation}

The generation of caustic structure and the anomalous magnification 
pattern associated with it makes planetary masses orbiting stellar lenses 
easier to detect than isolated lensing planets.  Even so, most planetary 
light curves will be anomalous because the source passed near, but 
not across a caustic (Fig.~15).  
How is the projected planet-star separation $b$ 
and the planet-star\\ 

\vglue -4.25cm
\hglue -1cm
\epsfxsize=13cm\epsffile{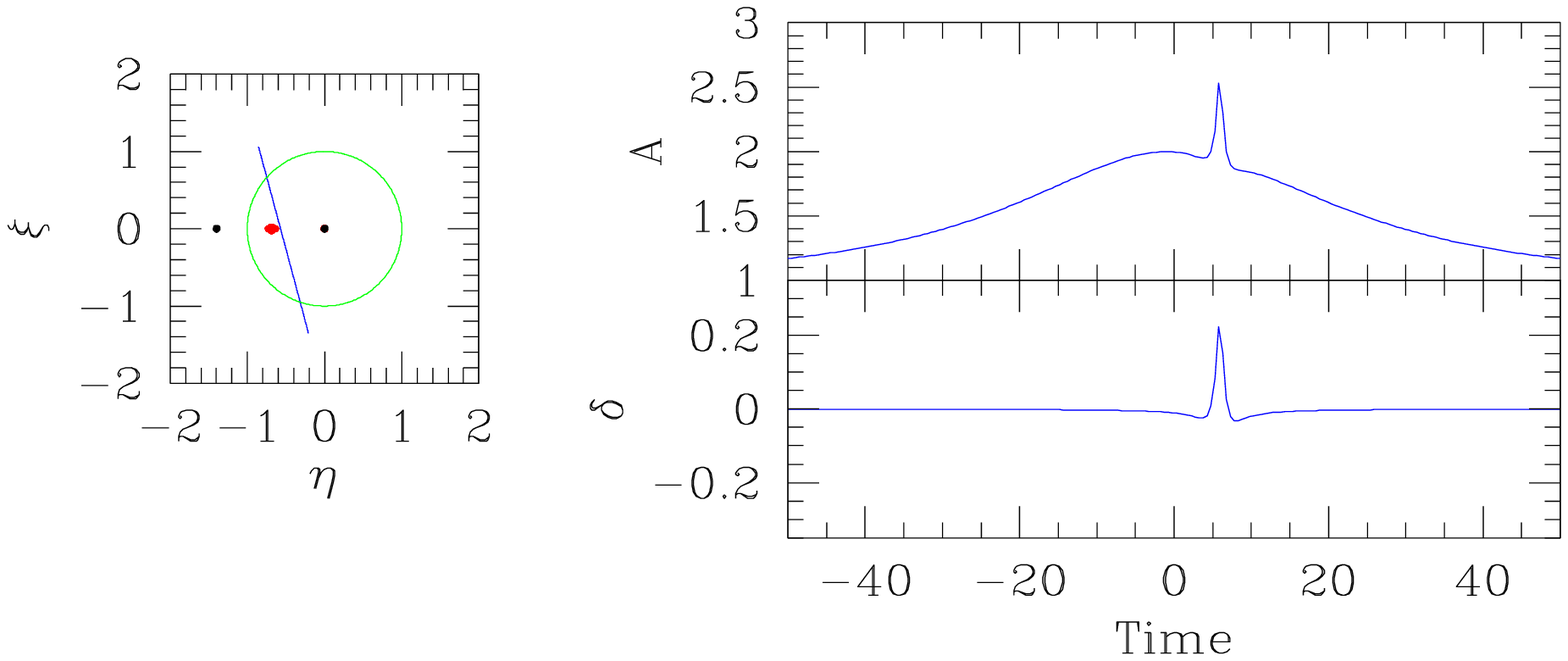}
\vglue -3.75cm{\small Fig.~15 --- 
{\bf Left:} A background point source travels along the (blue) trajectory 
that just misses the (red) caustic structure caused by a ``Jupiter'' with 
mass ratio $q=0.001$ located at 1.3 Einstein ring radii (several AU) 
from its parent stellar lens.  {\bf Right:}  The resulting light curve is 
shown in the top panel; the excess magnification $\delta$ associated 
with the planetary anomaly is shown in the bottom panel; time scale 
is in days.}

\noindent 
mass ratio $q = m_p/M_*$ extracted 
from a planetary anomaly atop an otherwise normal microlensing 
light curve?  In practice, the morphology of planetary light curve anomalies 
is quite complex, and detailed modeling of the excess magnification 
pattern (the anomalous change in the pattern due to the planet) 
is required, but the general principles can be easily understood.

\hglue -1.5cm
\epsfxsize=14cm\epsffile{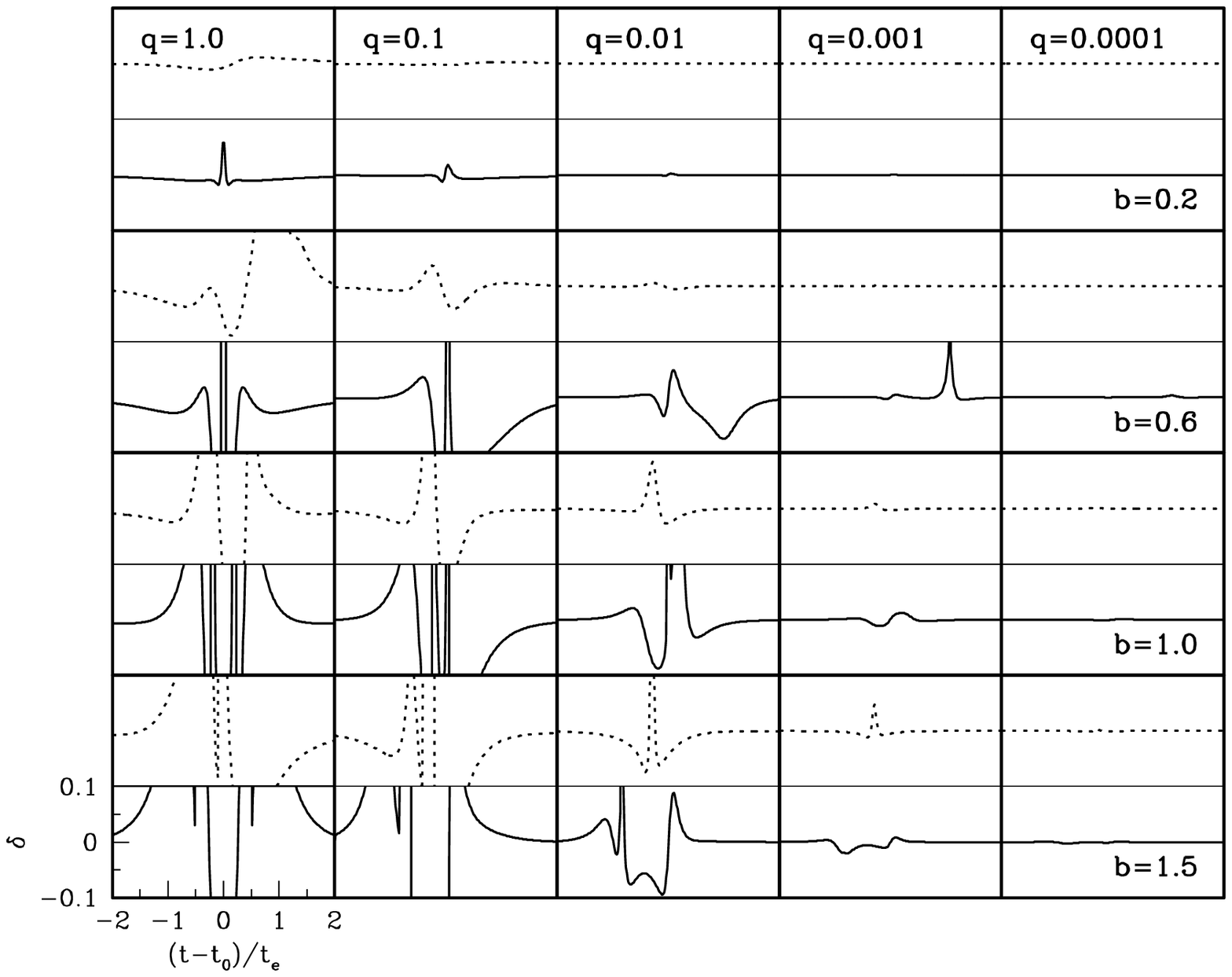}
\vglue -4.5cm{\small Fig.~16 --- 
Excess magnifications $\delta$ for the (solid and dotted) trajectories 
of Fig.~14 are shown for (the same) range of planetary mass ratios 
and projected separations.  ``Super-jupiters'' with $q \sim 0.01$ 
should create detectable anomalies for a significant fraction of 
source trajectories in high quality light curves.  (Adapted from 
Gaudi \& Sackett 1998.)}
\vskip 0.4cm

Since the planet and parent star lenses are 
at the same distance $D_L$ and travel across the line of sight 
with the same velocity $v_\perp$ (ignoring the planet's orbital motion), 
Eq.~16 shows that 
the mass ratio $q$ is equal to the square of the ratio of the Einstein ring 
radii $(\theta_p/\theta_E)^2$.  Observationally this can 
be estimated very roughly by the square of the ratio of the  
planetary anomaly duration to the primary event duration, $(t_p/t_E)^2$.  
The time difference between the primary and anomalous 
peaks (normalized to the Einstein time) gives an indication of the 
placement of the caustic structure within the Einstein ring and thus 
the position of the planet relative to the primary lens, $b$. 
The amplitude of the anomaly $\delta \equiv (A - A_0)/A_0$, where 
$A_0$ is the unperturbed amplitude, 
indicates the closest approach to the caustic 
structure and, together with the temporal placement of the anomaly,  
yields the source trajectory angle through the magnification pattern. 

Since the magnification pattern associated with planetary caustics 
for $b > 1$ and $b < 1$ planets is qualitatively different, detailed  
dense monitoring should resolve 
any ambiguity in the planetary position.
Light curve anomalies associated with $b > 1$ planets, like the one 
in Fig.~15, will have relatively large central peaks in $\delta$ surrounded by 
shallow valleys; $b < 1$ anomalies will generally have more rapidly 
varying alterations of positive and negative excess magnification, 
though individual exceptions can certainly be found.
 
From the shape of light curve anomalies alone, the mass of the 
planet is determined as a fraction of the primary lens 
mass; its instantaneous projected separation is determined 
as a fraction of the primary Einstein radius.  
Reasonable assumptions about the kinematics, distribution, and 
masses of the primary stellar lenses, together with measurements of the  
primary event duration $2 t_E$ and fraction of blended light 
from the lens should allow $r_E$ and $M_*$  
to be determined to within a factor $\sim3 - 5$.  
Detailed measurements of the planetary anomaly would then yield   
the absolute projected separation and planetary mass to about 
the same precision.   

\subsubsection{Durations and Amplitudes of Planetary Anomalies}

It is clear from Figs.~14 and 16 that, depending on the source 
trajectory, a range of anomaly durations $t_p$ and 
amplitudes $\delta$ are possible for a planetary 
system of given $q$ and $b$ (see also Wambsganss 1977).  
Nevertheless, rough scaling relations 
can be developed to estimate the time scales and amplitudes that 
will determine the photometric sampling rate and precision required 
for reasonable detection efficiencies to microlensing planetary systems.

For small mass ratios $q$, 
the region of excess magnification associated with 
the planetary caustic is a long, roughly linear region with a width 
approximately equal to the Einstein ring of the planet, $\theta_p$, and 
a length along the planet-lens axis several times larger.  
Since $\theta_p = \sqrt{q} \, \theta_E$,  
both the time scale of the duration and the cross section 
presented to a (small) source vary linearly with $\theta_p/\theta_E$ 
and thus with $\sqrt{q}$. 
Assuming a typical $t_E = 20$ days, the duration of the planetary 
anomaly is given roughly by the time to cross the planetary 
Einstein diameter, $2 \, \theta_p$, 

\begin{equation}
{\rm planet~anomaly~duration~} = 2 \, t_p \approx {\rm 1.7 \, hrs} \, 
                               (m/ \mearth)^{1/2} (M/ \msolar)^{-1/2}. 
\end{equation}

Caustic crossings can occur for any planetary mass ratio and should be 
easy to detect as long as the temporal sampling is well matched to 
the time scales above.  Most anomalies, however, will be more gentle 
perturbations associated with crossing lower amplitude excess 
magnification contours.   
At the most favorable 
projected lens-planet separation of $b=1.3$, and the most ideal 
lens location (halfway to the Galactic Center), well-sampled 
observations able to detect 5\% perturbations in the light curve 
would have planet sensitivities given roughly by (Gould \& Loeb 1992):

\begin{equation}
{\rm ideal~detection~sensitivity~} \approx 1\%~ 
                                   (m/ \mearth)^{1/2} (M/ \msolar)^{-1/2}
\end{equation}

\noindent
This ideal sensitivity is relevant only for planets at $b=1.3$; at the 
edges of the lensing zone the probabilities are about halved. 
Detection with this sensitivity requires photometry at the 1\% 
level, well-sampled over the duration of the planetary event.  

\vglue -0.5cm
\hglue 0.25cm
\epsfxsize=11.5cm\epsffile{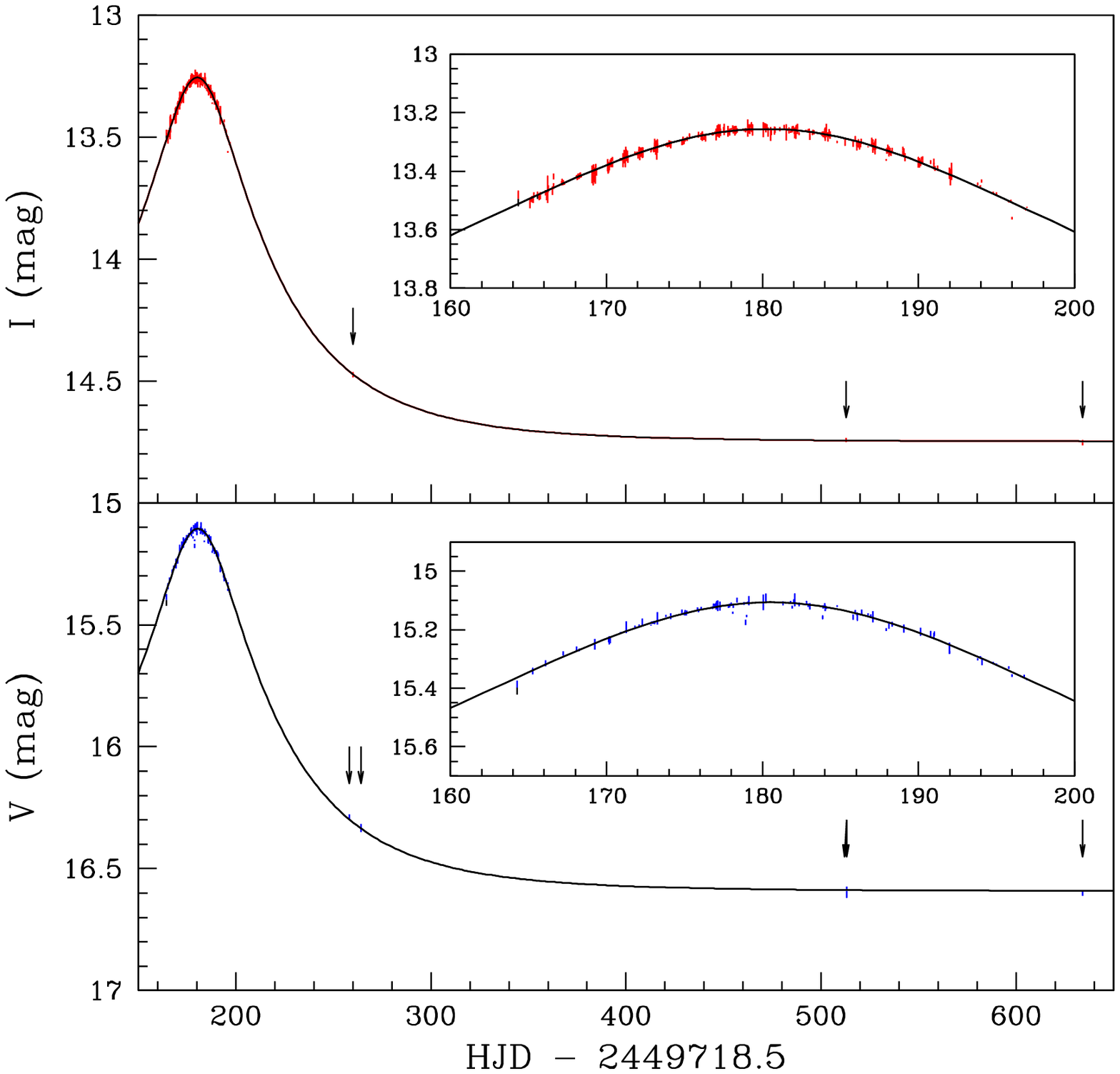}

\vglue -1cm{\small Fig.~17 --- PLANET collaboration 
monitoring of MACHO-BLG-95-13 in the I (upper) and V (lower) bands. 
Insets show a zoom around the peak of the event; arrows indicate 
points taken many months to more than a year later. Vertical scale is 
magnitudes; horizontal scale is days (Albrow \etal\ 1998a).}
\vskip 0.4cm

Can such photometric precision and temporal 
sampling be obtained in the crowded fields 
of the Galactic bulge where nearly all microlensing events are discovered?
Fig.~17 shows observations of one bright microlensing event 
monitored by the PLANET collaboration during its month-long pilot 
season in 1995 (Albrow \etal\ 1998a).  
The residuals from the single point-lens/point source 
light curve are less than 1\% for this event, and the typical 
sampling rate is on the order of once every 1-2 hours, even 
accounting for unavoidable longitudinal and weather gaps.

A true calculation of detection probabilities must integrate over the 
projected orbital separations and the distribution of 
lenses along the line of sight, must take into account 
the actual distribution of source trajectories probed by 
a particular set of observations, and the effect of uneven temporal 
sampling and photometric precision (Gaudi \& Sackett 1998).  
In the following section we 
discuss the additional complication of finite source effects that  
is encountered for very small mass planets for which the size of 
the planetary Einstein ring is comparable to or smaller than the source size,  
$\theta_p \ltorder \theta_*$.  

\subsection{Observational Rewards and Challenges}

What can be learned by observing a planetary anomaly in a microlensing 
light curve?  The duration, temporal placement relative to the event 
peak, and relative amplitude of the anomaly can be used to determine 
the mass ratio $q$ of the planetary companion to the primary (presumably 
stellar) lens and their projected angular separation $b$ in units of 
the Einstein ring radius $\theta_E$.
Since in general the lens will be far too distant to type spectrally 
against the bright background source (except possibly with 
very large apertures, see Mao, Reetz \& Lennon 1998), the 
absolute mass and separation must be determined statistically by 
fitting the properties of an ensemble of events with reasonable 
Galactic assumptions.  Measurements of other sorts of microlensing 
anomalies associated with source resolution, observer motion, or 
lens blending can produce additional constraints on the lens properties 
and thus on the absolute planetary characteristics.

Except for very large $a$ planets orbiting in nearly face-on 
($i \approx 0$) orbits, cooperative lensing effects between the lens and companion boost the detectability of lensing planets over that expected 
for planets in isolation. Since current detection and monitoring schemes focus 
on those events with an essentially random distribution of impact parameters 
$u_{min}$ for $u_{min} < 1$, the anomaly sensitivity is primarily 
restricted to planets in the ``lensing zone'' with projected separations of 
0.6 --- 1.6 times the Einstein ring radii of the primary lens.  
For typical distributions of lens masses, and lens and source distances, 
this translates into the highest probabilities for planets with 
instantaneous orbital separations {\it projected onto the sky plane\/} of 
$a_p \approx 5 \,$AU, with a zone of reduced detectability extending to 
higher $a_p$.  Since the efficiency of planetary detection in these zones 
is likely to be a few or a few tens of percent (Eq.~21), many  
microlensing events must be monitored with $\sim$1\% photometry on 
the $\sim$hourly time scales (Eq.~20) to achieve statistically meaningful 
constraints on the number planets in the Milky Way, and their distribution 
in mass and orbital radius.

What limits the photometric precision and temporal sampling? 
Round-the-clock monitoring of events, necessary for maximum sensitivity 
to the 1 -- 2 day durations of Jupiter-mass anomalies requires telescopes 
scattered in geographical longitude, at the South Pole, or in space.  
Temporal sampling is limited by the number of 
events monitored at any given time, their brightness, and the desired level of 
photometric precision.  Higher signal-to-noise can generally be 
obtained for brighter stars in less exposure time, but 
ultimately, in the crowded fields that typify any 
microlensing experiment, photometric precision is limited by 
confusion from neighboring sources, not photon statistics.  Pushing 
below $\sim$1\% relative photometry with current techniques 
has proven very difficult in crowded fields.

If an anomaly is detected, it must be distinguished from other 
intrinsic effects that could be confused with a lensing planet.  
Stellar pulsation on daily to sub-daily time scales in giant 
and sub-giant bulge stars is unlikely, 
but this and any form of regular variability 
would easily be recognized as periodic (\ie\ non-microlensing) 
with the dense and precise 
sampling that is required in microlensing monitoring designed to 
detect planets.  Star-spot activity may be non-negligible in giants, 
but will have a time scale characteristic of the rotation period 
of giants, and thus much longer than typical planetary anomalies.  
In faint dwarf stars spotting activity produces flux changes below 
the photometric precision of current experiments.  Flare activity 
should not be significant for giants and is expected to be chromatic, 
whereas the microlensing signal will always be achromatic 
(except in the case of source resolution by exceedingly low-mass 
planets).  

Blending (complete photometric confusion) by random, 
unrelated stars along the line-of-sight can dilute the apparent 
amplitude of the primary lensing event.  This will have some effect 
on the detection efficiencies, but most significantly 
--- with data of insufficient sampling and photometric 
precision --- will lead to underestimates for the time scale  
$t_E$ and impact parameter $u_{min}$ of the primary event, and thus also to 
mis-estimates of the planetary mass ratio $q$ and projected separation $b$. 

\subsubsection{Pushing the Limits: Earth-mass and Multiple Planets}

Planets with very small mass ratio will have caustic structure 
smaller than the angular size of typical bulge giants.  The ensuing 
dilution of the anomaly by finite source effects will present a large,  
but perhaps not insurmountable, challenge to pushing the microlensing 
planet detection technique into the regime of terrestrial-mass planets 
(Peale 1997, Sackett 1997).

Near small-mass planetary caustics, 
different parts of the source simultaneously cross regions 
of significantly different excess magnification;  
an integration over source area is required in order to derive the 
total magnification.  The severity of the effect can be seen in 
Fig.~18.  Earth-mass caustic crossings against even the smaller-radii bulge 
sources will present a challenge to current photometry in 
crowded fields, which is generally limited by seeing to stars above 
the turn-off mass.

The most numerous, bright microlensing sources in the bulge 
are clump giants with radii  
about 13 times that of the Sun (13 $R_\solar$), and thus angular radii    
of 7.6 microarcseconds ($\mu$as) at 8 kpc.  
Since a Jupiter-mass planet with $m_p = 10^{-3} \msolar$ has an 
angular Einstein ring radius of 32 $\mu$as at 4 kpc and 19 $\mu$as at 6 kpc, 
its caustic structure is large compared to the size of the source.  
An Earth-mass 
planet with $m = 3 \times 10^{-6} \msolar$, on the other hand, 
has an angular Einstein ring radius of 1.7 $\mu$as at 4 kpc and 
1 $\mu$as at 6 kpc, and will thus suffer slight finite source effects 
even against turn-off stars (1.7 $\mu$as), though 
the effect will be greatly reduced compared to giant sources 
(Bennett \& Rhie 1996). 

\vglue -0.75cm
\hglue -0.75cm
\epsfxsize=7.5cm\epsffile{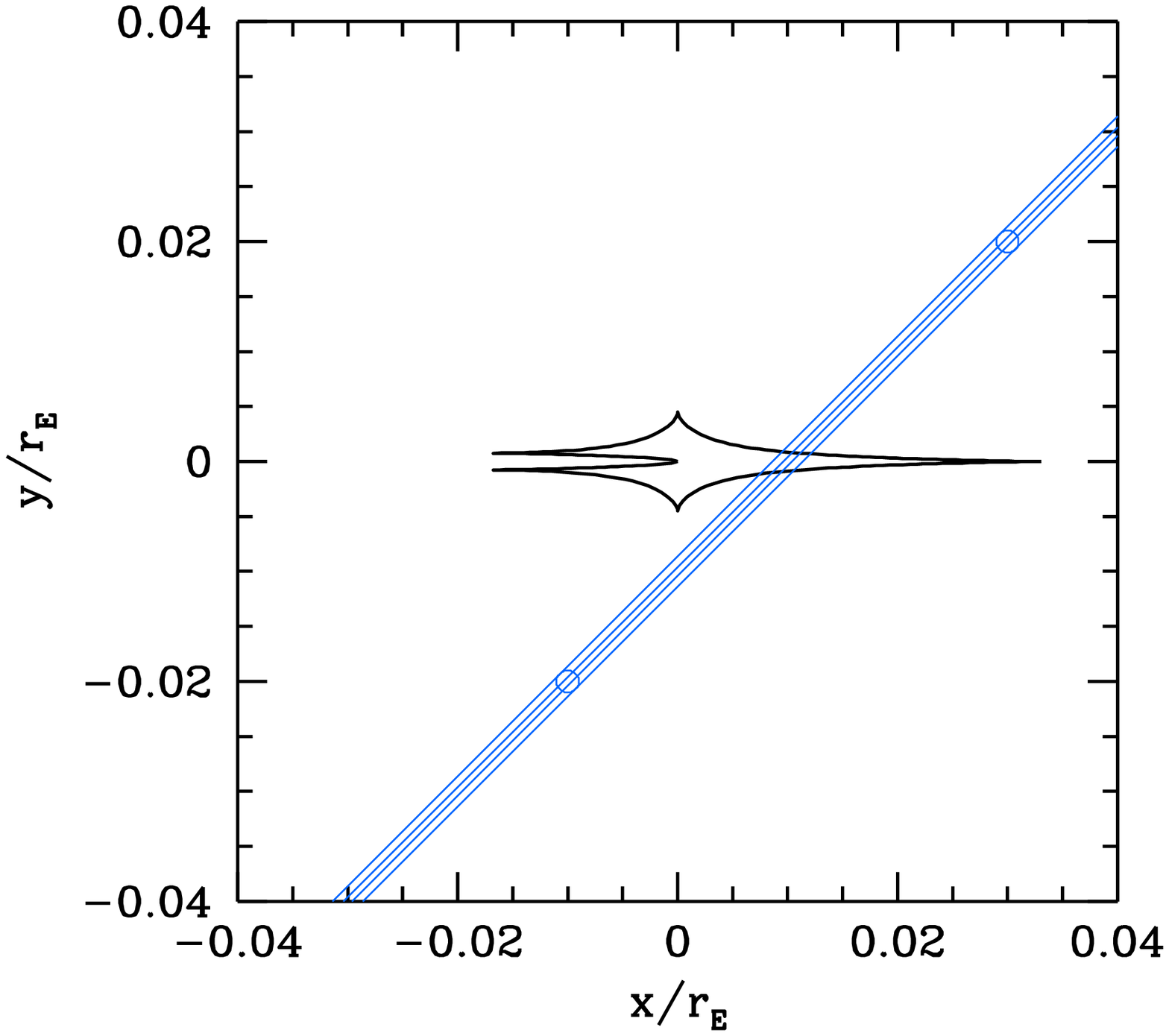}

\vglue -7.5cm
\hglue 5.5cm
\epsfxsize=7.5cm\epsffile{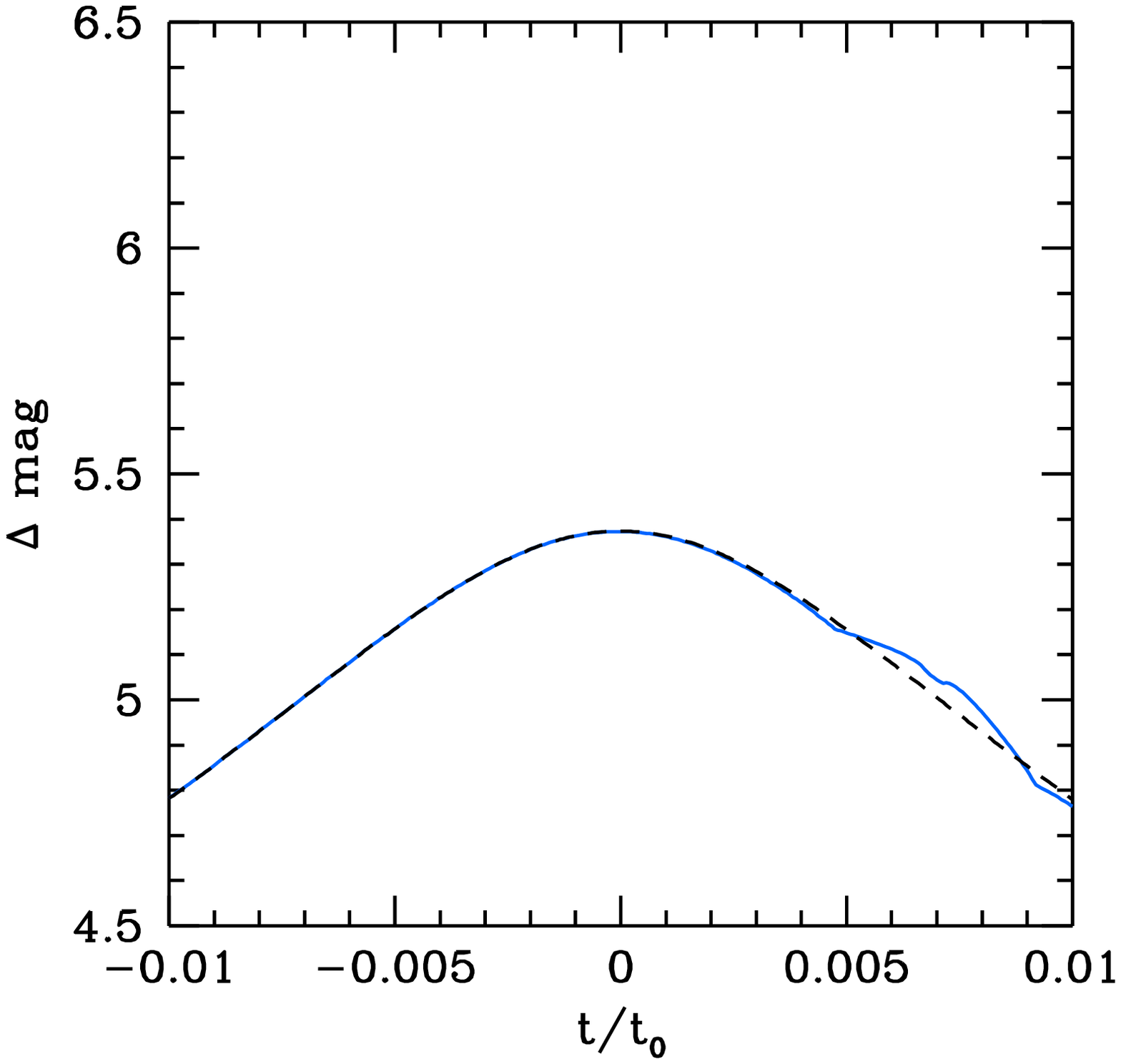}

\vglue -0.75cm{\small Fig.~18 --- {\bf Left:}  A source of angular 
size $\theta_* = 0.001 \, \theta_E$, typical of turn-off stars in the bulge, 
crosses the central caustic caused by a terrestrial-mass planet with 
mass ratio $q = 10^{-5}$.  {\it Right:}  Due to source resolution effects, 
the resulting anomaly differs from single-lens microlensing only 
at the $\sim$1\% -- 3\% level. Note the expanded spatial and temporal 
scales.  (Adapted from Paczy\'nski 1996.)}
\vskip 0.4cm

For extreme source resolution, in which the entire planetary caustic 
lies inside the projected source radius, the {\it excess fractional\/} 
magnification associated with the planetary anomaly scales with the square of 
the ratio of the planetary Einstein ring radius to the angular source size,  
$\delta \propto (\theta_p/\theta_*)^2$.  
On the other hand, source-resolved small $q$ anomalies will have 
longer durations than implied by Eq.~20, since the characteristic time scale  
is the time to cross the source $\theta_*$ (not $\theta_p$).  
Furthermore, the cross section for magnification at a given threshold now 
roughly scales with $\theta_*/\theta_E$ (not $\theta_p/\theta_E$), 
and is thus approximately independent of planetary mass.  

Because the anomaly amplitude is suppressed by source resolution, 
unless the photometry is excellent and continuous, small-mass 
planetary caustic crossings can be confused 
with large impact parameter large-mass planetary anomalies.  This 
degeneracy can be removed by performing multi-band observations 
(Gaudi \& Gould 1997): large impact parameter events 
will be achromatic, but sources resolved by small-mass caustics  
will have a chromatic signal due to source limb-darkening that is similar 
to (but of opposite sign from) that expected for planetary transits 
(\S2.3.1).  Source limb-darkening and chromaticity 
have now been observed during a caustic crossing of a stellar binary 
(Albrow \etal\ 1998b).

\vglue -0.5cm
\hglue 1.5cm
\epsfxsize=9.5cm\epsffile{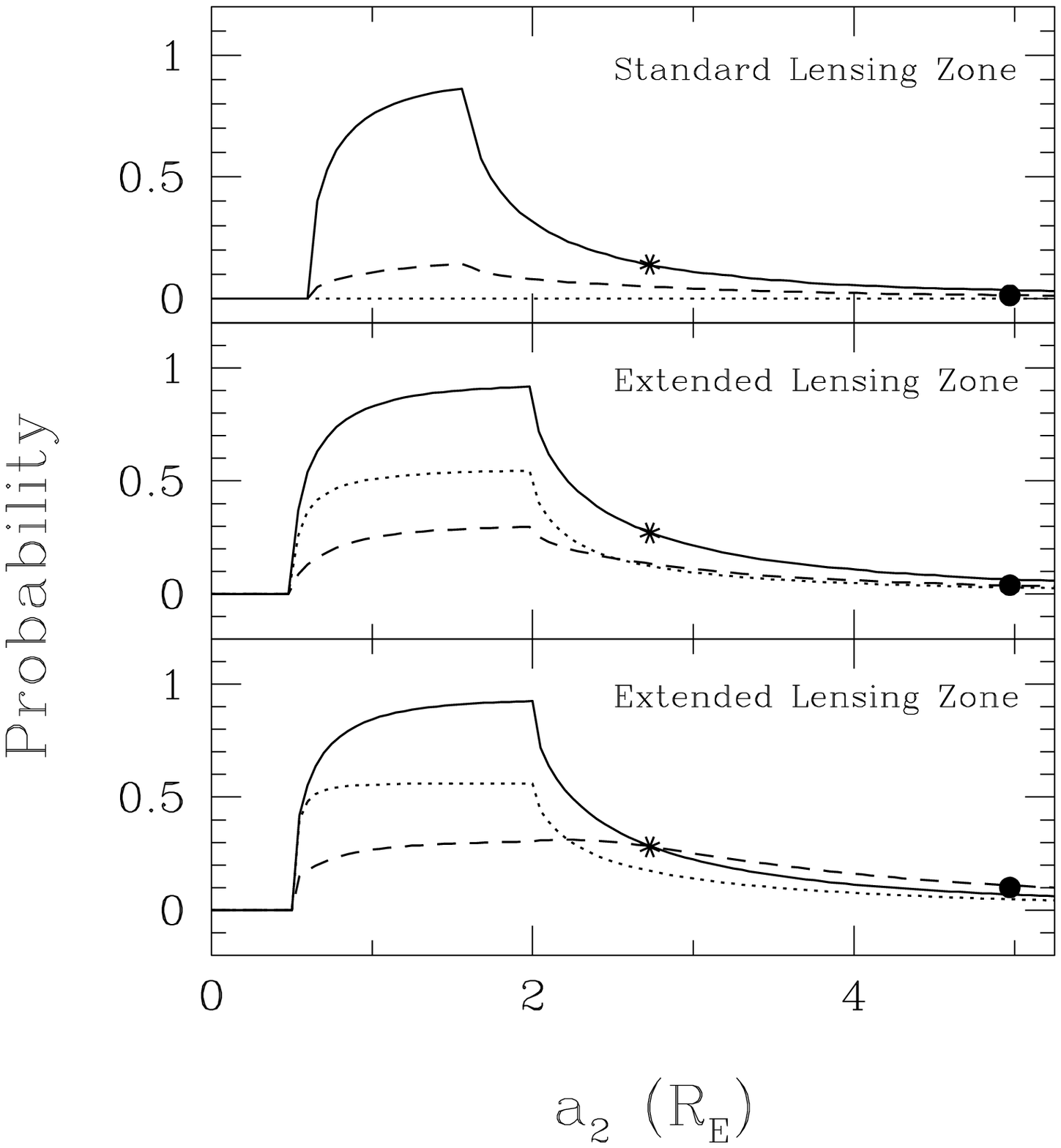}

\vglue -1.0cm{\small Fig.~19 --- {\bf Top:} Probability that 
two planets with true orbital radii $a_1$ and $a_2$ 
(in units of the Einstein ring $r_E$)
simultaneously have projected separations, 
$b_1$ and $b_2$, in the standard ``lensing
zone,'' defined as $0.6 < b < 1.6$.  The probability is shown
as function of $a_2$, for fixed $a_1=1.5$ (solid), 
$a_1=0.6$ (dotted) and $a_1=2.7$ (dashed).  
The probability for two planets with orbital radii of Jupiter and Saturn 
around solar-mass primary (star) and a $0.3M_{\odot}$ primary (dot) 
are shown.
{\bf Middle:} Same, but for the extended ``lensing zone,'' 
$0.5 < b < 2.0$.  {\bf Bottom:} The conditional probability that 
both $b_1$ and $b_2$ lie in the extended ``lensing
zone,'' given that either $b_1$ or $b_2$ satisfies this criterion. 
(Gaudi, Naber \& Sackett 1998.)}
\vskip 0.3cm

Finally, since all planetary lenses create a central caustic, low-impact 
parameter (high magnification) microlensing events that project the source 
close to the central caustic are especially efficient in producing 
detectable planetary anomalies (Griest \& Safizadeh 1998).  
For the same reason, however, the central caustic is affected by {\it all\/} 
planets in the system, and so --- if possible degeneracies due to the 
increased caustic complexity can be removed --- 
rare, low impact parameter events offer a promising way of simultaneously  
detecting multiple planets brought into or near the lensing zone by 
their orbital motion around the primary lens (Gaudi, Naber \& Sackett 1998). 
As Fig.~19 demonstrates, the statistical probabilities are large 
that a Jupiter or 47UMa 
orbiting a solar-mass star (solid and dotted lines, respectively) will 
instantaneously share the lensing zone with other planets of  
orbital radii of several AU.  However, the light curves resulting from crossing 
a multiple-planet central caustic may be difficult to interpret since 
the caustic structure is so complicated (Fig.~20).

\vglue -0.1cm
\hglue 0.5cm
\epsfxsize=9.5cm\epsffile{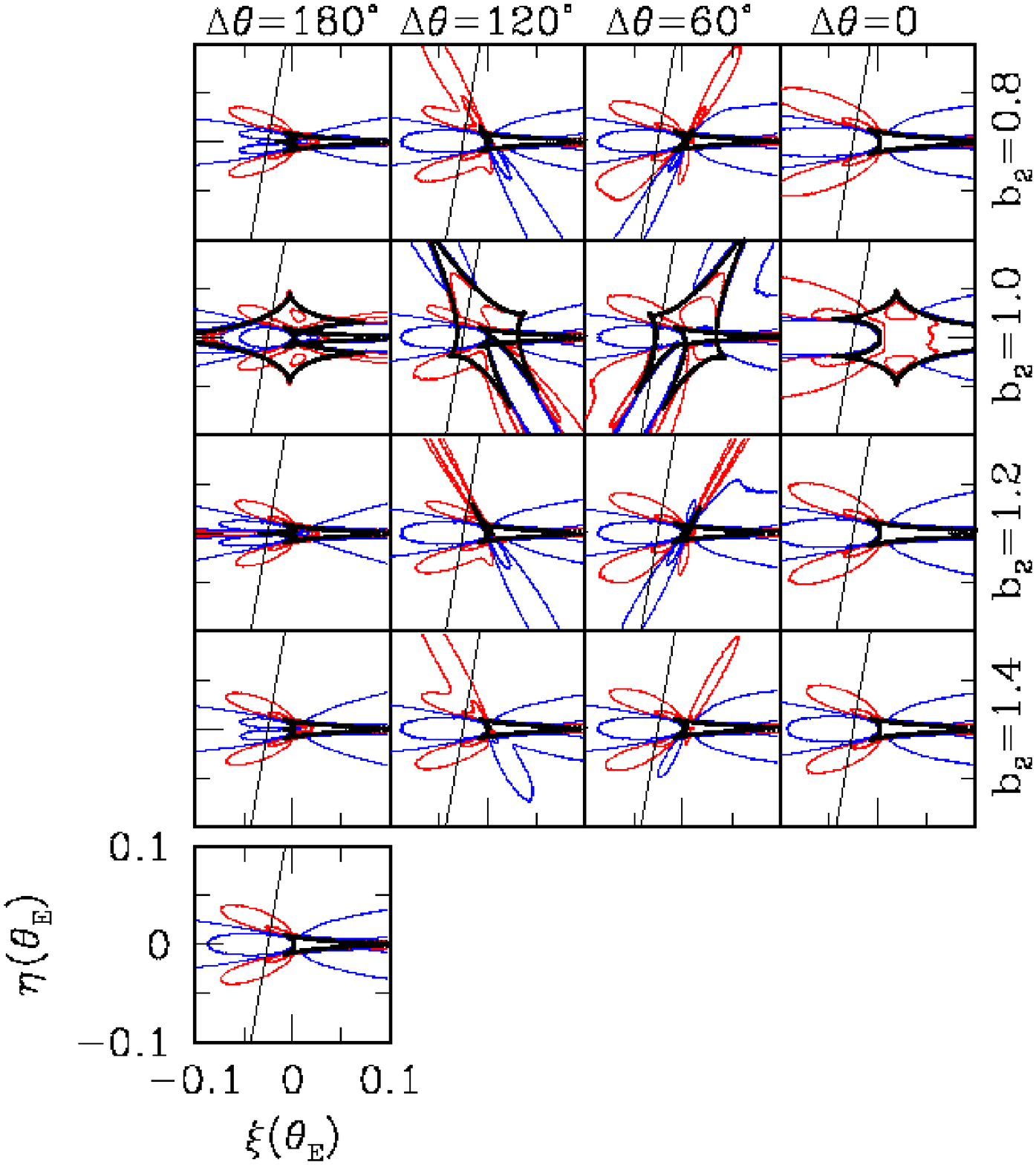} 

\vglue -0.25cm{\small Fig.~20 --- 
Contours of 5\% and 20\% fractional deviation $\delta$, 
as a function of source position in units of the $\theta_E$.  
The parameters of planet 1 are held fixed
at $q_1=0.003$, $b_1=1.2$; the projected separation 
$b_2$ and the angle between the axes, $\Delta\theta$, are varied
for a second planet with $q_2=0.001$.  Only planet 1 is present 
in the bottom offset panel.   Positive (red), negative (blue) and caustic 
($\delta=\infty$, thick line) contours are shown.
(Gaudi, Naber \& Sackett 1998.) }


\section{Photometric Mapping of Unseen Planetary Systems: \\
$~~~~~~$Matching the Tool to the Task}

Both the transit and microlensing techniques use frequent, high-precision 
monitoring of many stars to discover the presence of the 
unseen extra-solar planets.  The transit method monitors the light 
from the parent star in search of occultation by an unseen planet; 
the microlensing technique monitors light from an unrelated 
background source in search of lensing by an unseen planet orbiting 
an unseen parent star.  
Indeed, microlensing is the only extra-solar planetary search technique 
that requires {\it no photons from either the planet or the parent star\/} 
and for this reason is the method most sensitive to the  
distant planetary systems in our Galaxy.

The two techniques are complementary, both in terms of the information 
that they provide about discovered systems, and in terms of the 
range of planetary system parameters to which they are most sensitive.  
Multiple transit measurements of the same planet will yield its planetary 
radius $R_p$ and orbital radius $a$.  Characterization of a microlensing 
planetary anomaly gives the mass ratio $q \equiv m_p/M_*$ 
of the planet to lensing star 
and their projected separation $b$ at the time of the anomaly in units 
of $\theta_E$.  

\hglue 2.0cm
\epsfxsize=8cm\epsffile{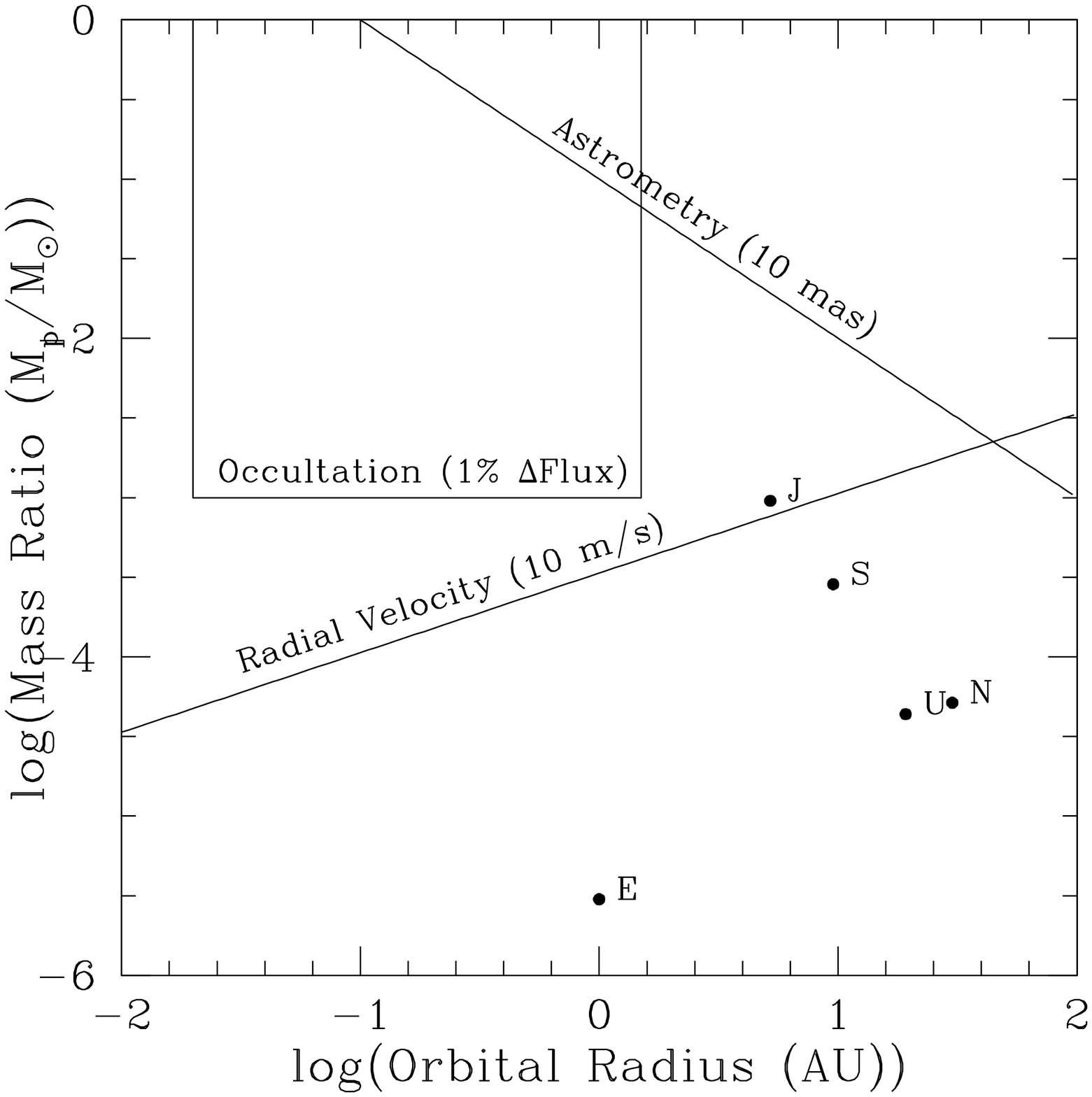}

\vglue -0cm{\small Fig.~21 --- Current detection thresholds 
for long-running programs that rely on planet orbital 
motion, shown as a function of planetary mass ratio and orbital radius.  
The occultation threshold must be multiplied by the 
appropriate geometric probability of a transit to derive detection 
efficiencies. 
Selected Solar System planets are shown.}
\vskip 0.4cm

Current ground-based photometry is sensitive to jovian-size 
occultations; space-based missions may extend this into the 
terrestrial regime.  The transit method is sensitive 
to planets with small $a$ because they create a detectable 
transit over a wider range of inclinations 
of their orbital planes,  and because they transit more often 
within a typical 5-year experiment.  These constraints limit the 
range of orbital radii to about $0.02 \ltorder a \ltorder 1.5 \, $AU 
for jovian-size planets.  If improvements in photometric precision would 
allow the detection of Earth-size planetary transits, this range would 
still be possible, but would suffer from noise due to star spot 
activity on time scales that could be confused with transits by small-mass 
planets with $a \ltorder 0.3\, {\rm AU}$.  

Since the planets in
own solar system fall roughly on the same log$ \, m_p$-log$ \, R_p$ 
relationship, it is reasonable to assume that jovian-size planets 
may also have jovian masses.  This assumption was used to place 
the current transit detection capability on the same plot (Fig.~21) 
with the current detection thresholds for radial velocity 
and astrometric techniques.  All three of these techniques require  
long-term projects to detect long-period (large $a$) planets since 
the measurements of velocity, position, or flux must be collected 
over at least one full orbital period.  The occultation threshold 
must be convolved with the geometric transit probability to 
derive efficiencies {\it per observed star\/}. 

Photometric precision in crowded fields together with 
source resolution effects limit current microlensing planet searches 
to Neptune masses and above.   The actual efficiency with which 
a given planetary system can be detected depends on its mass ratio 
and projected orbital separation.  Fig.~22 shows estimates of 
microlensing detection efficiency contours for planets of a given mass 
ratio and true orbital separation (in units of the Einstein radius).  
The contours are based on the work of Gaudi \& Gould (1997) for 
high-mass ratios, Gould \& Loeb (1992) and 
Gaudi \& Sackett (1998) for intermediate mass ratios,  
and Bennett \& Rhie (1996) for small ratios.  Integrations have 
been performed over the unknown but presumably randomly oriented 
inclinations and orbital phases.  
Although planets with $a$ in 
the lensing zone and orbiting in the sky plane 
are the easiest to detect, a tail of sensitivity extends to 
larger $a$ as well because inclination effects will bring large-$a$ planets  
into the projected lensing zone for some phases of 
their orbits.   The efficiencies assume $\approx$1\% photometry 
well-sampled over the post-alert part of the microlensing light curve.

Examination of Fig.~22 makes it clear that different indirect 
planetary 
search techniques will are sensitive to different portions of the 
log$ \, m_p$-log$ \, a$ domain.  Current ground-based 
capabilities favor the radial velocity method for short-period 
($a \ltorder 3$~AU) planets (see also Queloz, this proceedings).  
The occultation method will help populate the short-period part 
of the diagram, and if the programs are carried into space, will 
begin to probe the regime of terrestrial-sized planets in terrestrial 
environments.
Ground-based astrometry is favored for very 
long-period ($a \gtorder 40$~AU) planets, 
although the time scales for detection and confirmation are 
then on the order of decades.   Space-based astrometry promises 
to make this method substantially more efficient, perhaps by a factor 
of 100 (see also Colavita, this proceedings).  
Microlensing is the only technique 
capable of detecting in the near term substantial numbers of 
intermediate-period ($3 \ltorder a \ltorder 12$~AU) planets.  
Somewhat longer period planets may also be discovered by  
microlensing survey projects as independent ``repeating'' events 
in which the primary lens and distant planet act as independent lenses 
(DiStefano \& Scalzo 1999).  
Very short-period planets interior to 0.1~AU may be detectable using the 
light echo technique (Bromley 1992, Seager \& Sasselov 1998, 
Charbonneau, Jhu \& Noyes 1998), at least for parent 
stars with substantial flare activity, such as late M dwarfs.

\vglue -0.75cm
\hglue 1.5cm
\epsfxsize=10cm\epsffile{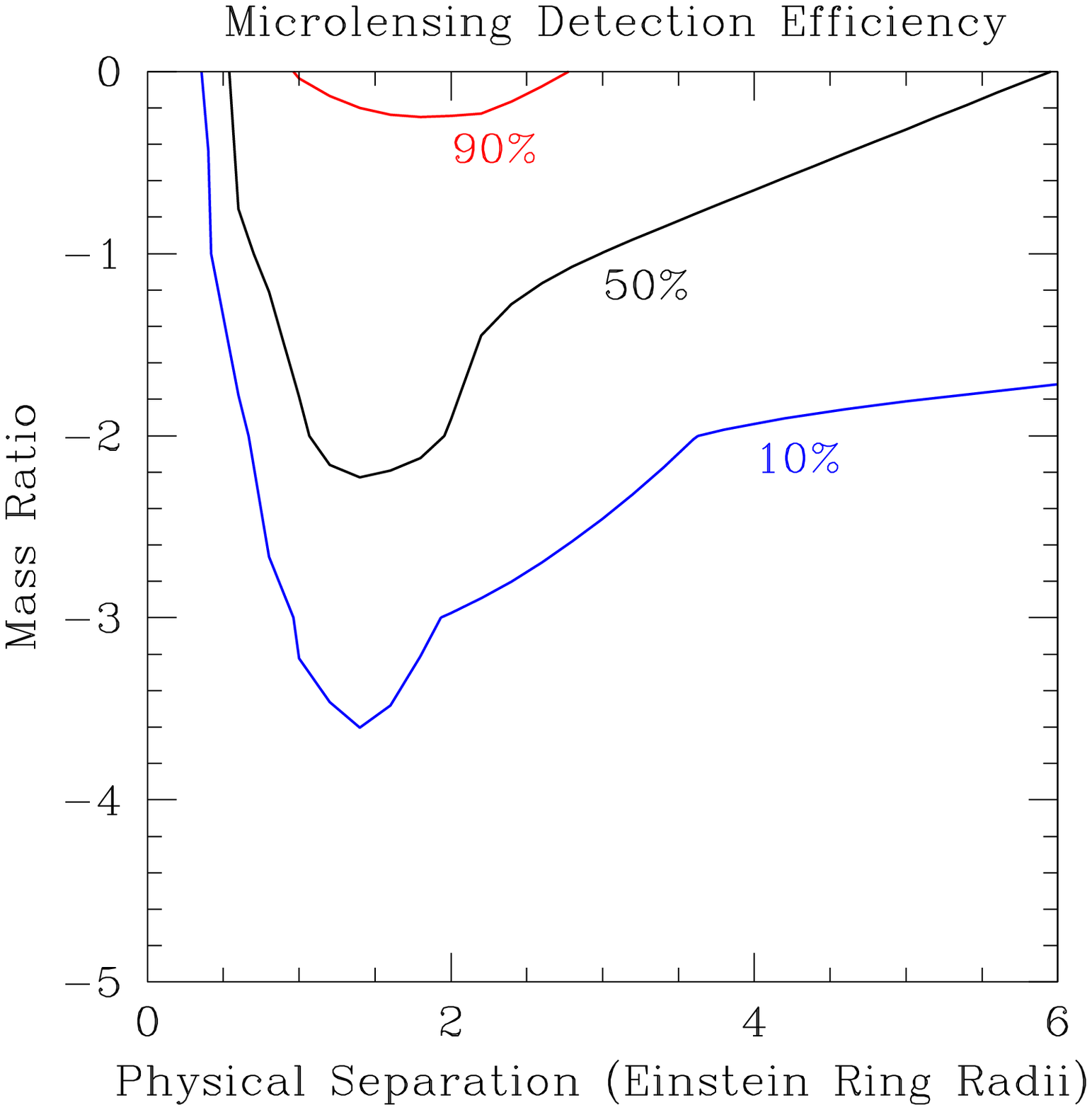}

\vglue -8.66cm
\hglue 2.2cm
\epsfxsize=7.85cm\epsffile{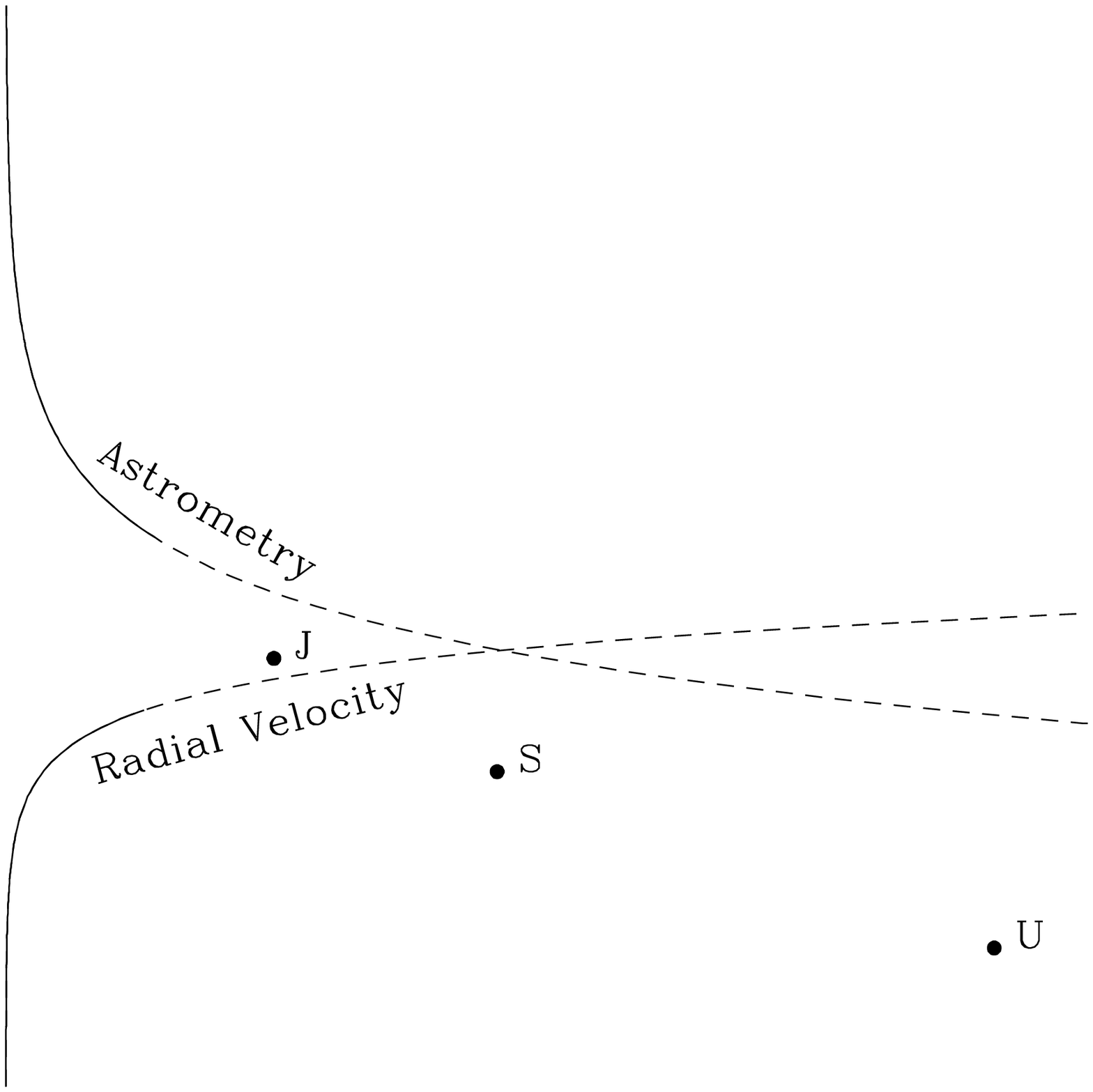}

\vglue 0.25cm{\small Fig.~22 --- Estimated detection efficiency 
contours for microlensing planet searches as a function of the logarithm 
of the planetary mass ratio $q \equiv m_p/M_*$ and the true orbital 
separation $a$ in units of the Einstein ring radius.  Efficiencies 
have been integrated over the phase and inclination of the orbits, 
under the assumption that they are circular.  
To make comparisons with other techniques, the Einstein ring radius 
is taken to be 3.5~AU.  Solid lines indicate 
what can be achieved in an observational program of 5-years 
duration or less.  Note that the vertical scale remains logarithmic, 
but the horizontal scale is now linear.  }
\vskip 0.4cm

The techniques are complementary in another sense as well. 
Those that rely on light from the parent star will be limited to 
nearby planetary systems, but will benefit from the ability to 
do later follow-up studies, including spectroscopy and interferometry.  
Microlensing, on the other hand, will see the evidence of a given 
planetary system only once, but can probe planetary frequency in 
distant parts of the Milky Way, and can collect statistics over a wide range 
of orbital separations in a relatively short time.
\vskip 0.25cm

{\small
\begin{center}
\begin{tabular}{l c c}
\noalign{\medskip\hrule\smallskip}
\multicolumn{3}{c}{TABLE II.  Comparison of Current Ground-Based Capabilities}\\
\noalign{\hrule}
\noalign{\smallskip\hrule\bigskip}
\medskip
~~~~  		          		&  OCCULTATION &  MICROLENSING \\

Parameters Determined	 		& $R_p$, $a$, $i$ 
					& $q \equiv m_p/M_*$,		\\
					& & $b \equiv a_p/R_E$		\\
Photometric Precision of 		& 0.1\%		& 1\%		\\
~~~Limits $R_p$ or $m_p$ to       	& Neptune 	& Neptune \\
Orbital Radius Sensitivity &$\sim 0.02 - 1.5\,$AU 
					& $\sim 1 - 12\,$AU	\\
Typical Distance of Systems		& $ < 1\,$kpc	& $ 4 - 7\,$kpc \\
Number of Stars to be Monitored	      	& few $10^3$ 	& few $10^2$ \\	    ~~~for Meaningful Jovian Sensitivity at & $\ltorder$1 AU &$\sim$5 AU \\
					&		&		\\
{\it In Principle Possible to Detect:\medskip}	&	&		\\
Multiple Planets			& yes		& yes		\\
Planets around Binary Parent Stars	& yes		& yes		\\
Earth-mass Planets in Future		& yes (space)   & yes 	\\
\noalign{\medskip\hrule\smallskip}
\end{tabular}
\end{center}
}

\subsection{Toward the Future}

The field of extra-solar planets is evolving rapidly.  
The number of groups conducting transit and microlensing planet 
searches, planning future programs, and providing theoretical support 
is growing at an ever-increasing rate.  
For that reason, this series of lectures has centered on the 
principles of the techniques rather than reviewing the current 
players.  In order to help the reader keep pace with this 
accelerating activity, however, a list of relevant Internet Resources 
with links to occultation and microlensing planet search groups 
is provided at the end of this section. 
What can we expect in the next decade from these research teams?  

Several ground-based transit searches are already underway 
(Villanova University, TEP, WASP, Vulcan, and EXPORT).
Some focus on high-quality 
photometry of known eclipsing binaries.  This is likely to increase 
the probability of transits --- if planets exist in such binary systems.  
Two transit-search teams recently issued 
(apparently contradictory) claims for a possible planet detection 
in the eclipsing M-dwarf system known as CM Draconis 
(IAU circulars 6864 and 6875, see also Deeg \etal\ 1998), but  
no clear, undisputed planetary signal has seen.  
One class of planets known to exist in reasonable numbers and also relatively 
easy to detect via occultation is the ``hot jupiter;'' the planet in 
51~Peg is a prototype of this class.  
If such a planet is the size of Jupiter, its orbital plane would have a  
$\sim$10\% chance of being sufficiently inclined to produce 
a detectable eclipse of a solar-type parent as seen from Earth.   
An aggressive ground-based program should be able to detect large 
numbers of such planets in the next decade --- planets that could be 
studied with the radial velocity technique thereby yielding both planetary 
mass and radius.
Space-based missions (COROT and KEPLER) planned for launch within 
this decade should have the sensitivity to detect transits from 
terrestrial-mass objects, but in order to detect Earth-like planets 
in Earth-like environments (\ie\ orbiting solar-type stars at 1~AU) they 
will need long mission times.

Microlensing planet searches are being conducted or aided by international 
collaborations (PLANET, GMAN, MPS, MOA, and EXPORT) that intensely monitor the 
events discovered by microlensing search teams (EROS, MACHO, OGLE, and MOA).  
MACHO and OGLE electronically alert on-going microlensing in the direction of 
the Galactic Bulge on a regular basis: at given time during the 
bulge observing season several events are in progress and 
available for monitoring.  Both the PLANET and GMAN collaborations 
have issued real-time secondary alerts of anomalous behavior (including 
binary lenses, source resolution, `` lensing parallax''), 
but to date no clear detection of a lensing planet has been announced.  
Especially if caustics are crossed, it may be possible to obtain 
additional information on microlensing planets from the sky  
motion of the caustics during the event that is induced by planetary 
motion (Dominik 1998).   
The number of high-quality microlensing light curves monitored 
by the PLANET collaboration is already beginning to approach that required  
for reasonable jovian detection sensitivities (Albrow \etal\ 1998a), 
so meaningful results 
on Jupiter look-alikes can be expected within the next few years.  

As more telescopes, more telescope time, and wider-field detectors are 
dedicated to dense, precise photometric monitoring capable of 
detecting planetary transits and planetary microlensing, we can feel certain  
that --- if jovian planets with orbital radii less than $\sim$6~AU 
exist in sufficient numbers --- they will 
be detected in the next few years by these techniques. 

\newpage


\vglue 0.1cm

\subsection*{Acknowledgments}

I am grateful to NATO for financial support and to the Institute's 
efficient and gracious scientific organizers, 
Danielle Alloin and (the sorely missed) Jean-Marie Mariotti, for 
a productive and pleasant school.  
It is also a pleasure to thank B. Scott Gaudi 
for assistance in the preparation of some of the figures in the 
microlensing section and for permission to 
show the results of our work before publication.


\vskip 0.5cm
\subsection*{INTERNET RESOURCES}

\vskip 0.3cm

\noindent
{\it General Extra-Solar Planet News:\/}

\vskip 0.2cm

\noindent
Extrasolar Planets Encyclopedia (maintained by J. Schneider):\\ 
http://www.obspm.fr/departement/darc/planets/encycl.html

\vskip 0.2cm
\noindent
and the mirror site in the U.S.A.:\\
http://cfa-www.harvard.edu/planets/

\vskip 0.4cm

\noindent
{\it Occultation:\/}

\vskip 0.3cm

\noindent
EXPORT:  http://pollux.ft.uam.es/export/

\vskip 0.2cm

\noindent
TEP:  http://www.iac.es/proyect/tep/tephome.html

\vskip 0.2cm

\noindent
Villanova University:  http://www.phy.vill.edu/astro/index.htm

\vskip 0.2cm

\noindent
VULCAN:  http://www.iac.es/proyect/tep/tephome.html

\vskip 0.2cm

\noindent
WASP:  http://www.psi.edu/~esquerdo/wasp/wasp.html

\vskip 0.4cm

\noindent
{\it Microlensing:\/}

\vskip 0.3cm

\noindent
EROS:  http://www.lal.in2p3.fr/EROS

\vskip 0.2cm

\noindent
MACHO:  http://wwwmacho.anu.macho.edu

\vskip 0.2cm

\noindent
MACHO Alert Page:  http://darkstar.astro.washington.edu

\vskip 0.2cm

\noindent
OGLE:  http://www.astrouw.edu.pl/$\sim$ftp/ogle

\vskip 0.2cm

\noindent
MOA:  http://www.phys.vuw.ac.nz/dept/projects/moa/index.html

\vskip 0.2cm

\noindent
MPS:  http://bustard.phys.nd.edu/MPS/

\vskip 0.2cm

\noindent
PLANET:  http://www.astro.rug.nl/$\sim$planet

\newpage


\end{document}